\documentclass[journal]{IEEEtran}
\usepackage{graphicx}
\usepackage{booktabs}
\usepackage{float}
\usepackage{amssymb,amsfonts}
\usepackage{amsmath,bm}
\usepackage{listings}
\usepackage{url}
\usepackage{algorithm,algorithmicx}
\usepackage{algpseudocode}
\usepackage{arydshln}
\usepackage{subfigure}
\usepackage{cite}
\usepackage{scalerel}
\usepackage{tikz}
\usepackage{hanging}
\usepackage[switch]{lineno}
\usepackage{blindtext}
\usetikzlibrary{svg.path}

\definecolor{orcidlogocol}{HTML}{A6CE39}
\tikzset{
  orcidlogo/.pic={
    \fill[orcidlogocol] svg{M256,128c0,70.7-57.3,128-128,128C57.3,256,0,198.7,0,128C0,57.3,57.3,0,128,0C198.7,0,256,57.3,256,128z};
    \fill[white] svg{M86.3,186.2H70.9V79.1h15.4v48.4V186.2z}
                 svg{M108.9,79.1h41.6c39.6,0,57,28.3,57,53.6c0,27.5-21.5,53.6-56.8,53.6h-41.8V79.1z M124.3,172.4h24.5c34.9,0,42.9-26.5,42.9-39.7c0-21.5-13.7-39.7-43.7-39.7h-23.7V172.4z}
                 svg{M88.7,56.8c0,5.5-4.5,10.1-10.1,10.1c-5.6,0-10.1-4.6-10.1-10.1c0-5.6,4.5-10.1,10.1-10.1C84.2,46.7,88.7,51.3,88.7,56.8z};
  }
}
\newcommand\orcidicon[1]{\href{https://orcid.org/#1}{\mbox{\scalerel*{
\begin{tikzpicture}[yscale=-1,transform shape]
\pic{orcidlogo};
\end{tikzpicture}
}{|}}}}

\usepackage{hyperref}
\usepackage{stfloats}

\newcommand{\tr}{{\rm{tr}} }
\newcommand{\true}{{\rm{true}} }
\newcommand{\vecc}{{\rm{vec}} }
\newcommand{\mat}{{\rm{mat}} }
\newcommand{\diag}{{\rm{diag}} }

\ifCLASSINFOpdf
\else
\fi
\begin{document}
%
\title{\Huge{Unified Attitude Determination Problem from Vector Observations and Hand-eye Measurements}}
%
%
%

\author{Jin Wu {$^{\orcidicon{0000-0001-5930-4170}\,}$},~\IEEEmembership{Member,~IEEE}
            
\thanks{The author is with Department of Electronic and Computer Engineering, Hong Kong University of Science and Technology, Hong Kong, China and is also with Tencent Robotics X. (e-mail: jin\_wu\_uestc@hotmail.com; chinajinwu@tencent.com)}
}

\maketitle
\begin{abstract}
The hand-eye measurements have recently been proven to be very efficient for spacecraft attitude determination relative to an ellipsoidal asteroid. However, recent method does not guarantee full attitude observability for all conditions. This paper refines this problem by taking the vector observations into account so that the accuracy and robustness of the spacecraft attitude estimation can be improved. The vector observations come from many sources including visual perspective geometry, optical navigation and point clouds that frequently occur in aerospace electronic systems. Completely closed-form solutions along with their uncertainty descriptions are presented for the proposed problem. Experiments using our simulated dataset and real-world spacecraft measurements from NASA dawn spacecraft verify the effectiveness and superiority of the derived solution.
\end{abstract}

\begin{IEEEkeywords}
Attitude Determination, Vector Observations, Hand-eye Measurements, Uncertainty Description, Optical Navigation
\end{IEEEkeywords}

%
\IEEEpeerreviewmaketitle

\section{Introduction}
\subsection{Background and Motivations}
\begin{figure}[ht]
\centering
\includegraphics[width=0.5\textwidth]{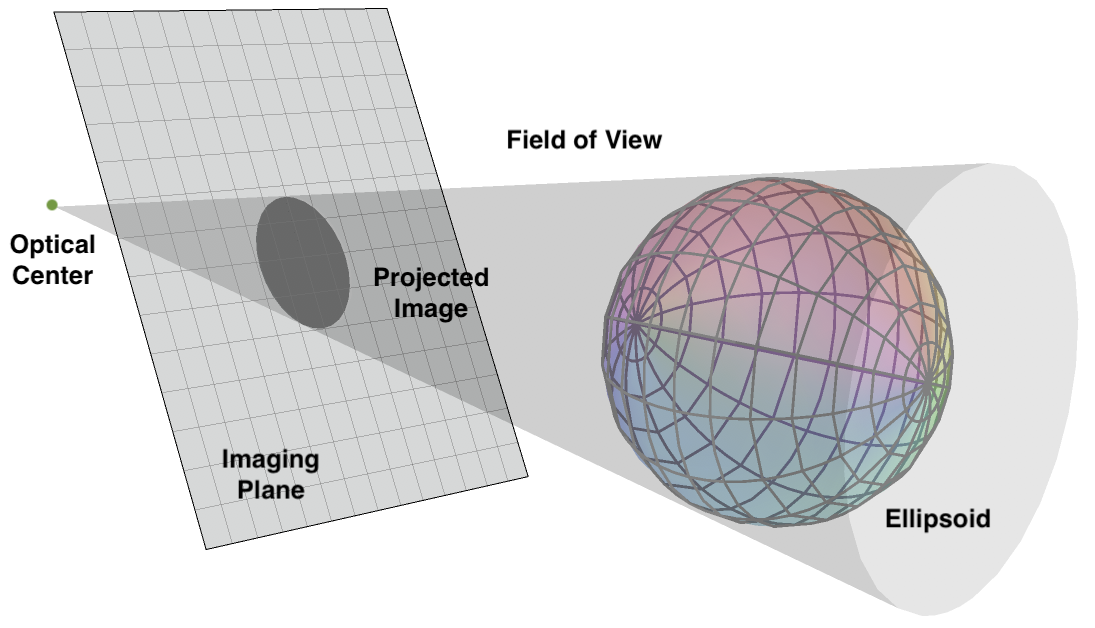}
\caption{The projection of an ellipsoid onto the imaging plane where the colors on the ellipsoid indicate some non-opportunistic features.}
\label{fig:chart}
\end{figure}

%
%
%
%
\IEEEPARstart{T}{he} research around planetary systems especially our solar system, has revealed many amazing secrets of the outer space. There are eight major planets and some other dwarf planets as well as asteroids in the solar system. In 1977, two Voyager spacecrafts were launched by National Aeronautics and Space Administration (NASA) to discover these celestial bodies \cite{kohlhase1977voyager}. Recently, some robots are sent to collect geology data from some planets and their satellites e.g. the Spirit launched in 2004 \cite{arvidson2008spirit} and the Curiosity launched in 2011 by NASA for discovery of the Mars \cite{hassler2014mars}, the YuTu (Jade Rabbit) launched in 2013 by China National Space Administration (CNSA) for inspection of the Moon \cite{xiao2015young}. A kernel problem behind these space projects is the precise navigation of various spacecrafts. The stability of nowdays space navigation systems, mainly constituted by strapdown installed measurement units, has been highly dependent on the accuracy of the attitude determination sub-system.\\
\indent Apart from the advances in inertial navigation technologies \cite{YuanxinWu2005,Wang2015,Wu2019}, attitude determination from heterogeneous sensor sources is a crucial problem in aerospace engineering since any spacecraft requires high-precision attitude information for motion control. The optimal attitude determination using vector observations from star trackers, sun sensors and magnetometers, namely the Wahba's problem, has been studied for over fifty years \cite{Crassidis2007}. Many efficient algorithms including the QUaternion ESTimator (QUEST, \cite{Shuster1981}), The Estimator of Optimal Quaternion (ESOQ, \cite{Mortari2000}), The Optimal Linear Attitude Estimator (OLAE, \cite{Mortari2007}) and etc. \cite{Choukroun2006} are developed to achieve accurate and reliable estimation. When the gyroscope measurements are taken into account, the vector-observation attitude estimation can be significantly enhanced via nonlinear observers on the special orthogonal group $SO(3)$ \cite{Mahony2008,Bahrami2017}. The hand-eye measurements formulate the attitude determination problem of the hand-eye type $\bm{AX} = \bm{XB}$ where $\bm{A}, \bm{B}$ are known and $\bm{X}$ is the unknown transformation to be figured out. The terminology 'hand-eye' was first introduced into robotics in late 1980s for calibration i.e. extrinsic transformation estimation between the robotic gripper and attached camera so that visual motion reconstruction can be precisely mapped to the gripper frame for accurate grasping \cite{Tsai1989,Daniilidis1999}. The idea of using the hand-eye measurements has recently been discovered by Modenini, who proposed a new method using images of ellipsoids. Modenini's method has been verified to be efficient with attitude accuracy of up to several arc seconds \cite{Modenini2018,Modenini2019}. Recently, an improved version with dimensionless matrices has also been proposed \cite{Modenini2019a}. Fig. \ref{fig:chart} shows such principle of imaging a space ellipsoidal object.\\

\begin{figure*}[ht]
\centering
\begin{minipage}[t]{2in}
\centering
\includegraphics[width=2in]{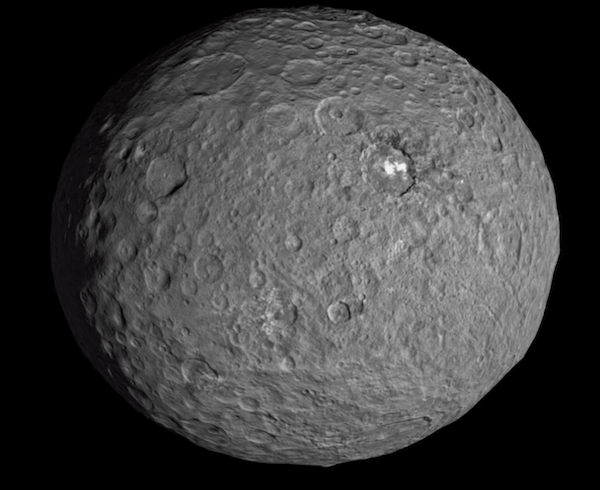}
\end{minipage}\begin{minipage}[t]{2in}
\centering
\includegraphics[width=2in]{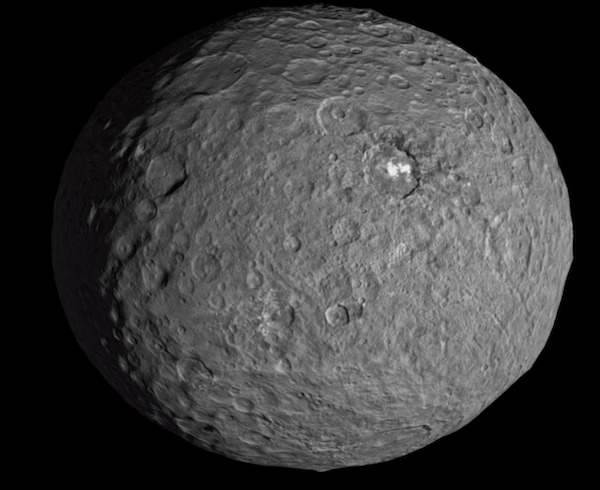}
\end{minipage}\begin{minipage}[t]{2in}
\centering
\includegraphics[width=2in]{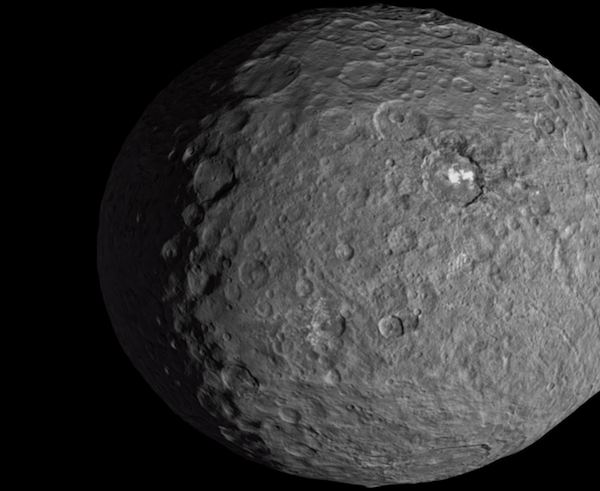}
\end{minipage}
\caption{The shading conics of the simulated Ceres with multiple light projection positions.}
\label{fig:conic}
\end{figure*}

\indent Modenini's new algorithm has given us a brand-new tool for spacecraft attitude determination relative to an ellipsoid. The problem studied in this paper is an extension of Modenini's contribution in which only one pair of symmetric hand-eye measurements are employed for attitude determination. According to the limitation of perspective geometry, any single shot of the celestial body can not completely reflect the exact details of 3D reconstruction \cite{man1982computational}. Therefore, the observability of attitude angles in all directions can not be fully guaranteed which requires further verification with historical estimates from other sources \cite{Modenini2018,Modenini2019}. To solve the problem, in this paper, instead, we incorporate the vector and hand-eye measurements together aiming for precise relative attitude determination between spacecraft and astroid. In Modenini's recent study, the hand-eye equation is established by means of the perspective geometry when the camera captures the image of an ellipsoid. A common knowledge exists in our mind is that there may be some distinctive (non-opportunistic) features on the surface of the ellipsoid, indicating that vector observations can form a unified attitude determination system together with hand-eye measurements. Another shortcoming of the existing results in \cite{Modenini2018,Modenini2019} is that the solving process depends on the spectrum decomposition which is nonlinear and an analytical covariance analysis can hardly be performed. The covariance information, however, guarantees the later fusion with other sensors using a Kalman filter, which is a common practice for reliability in aerospace engineering.
\subsection{Sources of Vector Observations}
There are many approaches that can provide vector observations. According to the principles of computer vision and robotics, the vector observations can be obtained either using point clouds from 3D laser \cite{Bercovici2017}, terrian reconstruction from the synthetic aperture radar (SAR) or rotation matrix from perspective-n-point (PnP, \cite{Gao2003}). Here we introduce three major ones.
\subsubsection{Perspective Geometry}
In aerospace engineering, practitioners usually first estimate the camera poses with respect to some celestial references. Then the camera poses can be transformed into the spacecraft attitude through pre-calibrated base-camera parameters. To achieve this, the perspective geometry is often invoked which consists of two representative methods i.e. the perspective-n-point (PnP) algorithm and epipolar geometry \cite{zhang1998determining}.\\
\indent The PnP solves the pose estimation problem when we have some interesting features in the 2D image and their corresponding perspective coordinates in the 3D frame. To get the accurate coordinates in the 3D frame, the precise 3D model or measurement of the imaged object is needed. The 2D-3D correspondences can be determined according to current 2D-3D registration techniques. The mathematical framework of the PnP problem allows for multiple solvers with various precisions including the direct linear transform (DLT, \cite{abdel2015direct}), efficient PnP (EPnP, \cite{Lepetit2009}), bundle adjustment (BA, \cite{Briskin2017}) and some recent analytical solutions \cite{Cai2018}.
\subsubsection{Optical Navigation}
The orbits of some celestial bodies can be acquired using historical celestial observations. With such data, the planets in the solar system and their relations with the Sun will be beneficial for spacecraft attitude determination. For instance, using conic extraction from binarized images of Moon, the Moon-Sun attitude sensor mechanism is established \cite{Mortari1997}. For aerospace engineering, when the orbit of the astroid is known and the luminance of the astroid is satisfactory for binarization, the vector observations can also come from such Asteriod-Sun relationship. Fig. \ref{fig:conic} depicts the conic variations of the modeled Ceres under different Sun projections.\\
\indent From another aspect, if the relative position between the Sun and the planet is known according to existing orbitary observations, then the Sun sensor, as a special kind of star trackers, can also provide observed Sun vector observation \cite{Choukroun2010}. 
\subsubsection{Point Clouds}
There are some airborne instruments that can give direct or indirect point-cloud measurements e.g. the laser scanner and SAR. By comparing the point clouds with existing 3D terrain, the spacecraft attitude is computed. When there is completely no knowledge of 3D terrain, the relative attitude can also be propagated sequentially using results from the iterative closest points (ICP, \cite{Besl1992,Wu2019symbolic}).

\subsection{Main Contributions and Arrangement of Contents}
The main contributions of this paper is to study the feasibility of such attitude determination scheme using vector observations and hand-eye measurements simultaneously. We are aimed to solve the following problems:
\begin{enumerate}
\item Derive the completely analytical solution to the proposed unified attitude determination problem.
\item Give an intuitive closed-form covariance analysis on the derived solution.
\item Study the characteristics of the proposed solution subject to different types of input values.
\item Study the real-world performances of the proposed scheme according to authentic spacecraft data.
\end{enumerate}

The remainder of this paper is organized as follows: Section II contains notation descriptions, problem formulation and proposed fully analytical solution. Covariance analysis is also given in the Section II showing some probabilistic characteristics of the proposed scheme. In Section III we conduct several experiments to deduce the in-flight attitude determination accuracy and robustness. Section IV consists of concluding remarks and future works.

\section{Problem Formulation and Solution}
\subsection{Mathematical Preliminaries}
We inherit the main usages of notations presented in \cite{Wu2019tim}. The $n$-dimensional Euclidean vector space is described with ${\mathbb{R}^n}$. We use $\mathbb{R}^{n \times m}$ to denote the real space containing all matrices with row dimension of $n$ and column dimension of $m$. The identity matrix has the notation of $\bm{I}$ and owns its proper size. ${{\bm{X}}^{\top}},{{\bm{X}}^{ - 1}}$ mean the transpose and inverse of a given matrix $\bm{X}$ respectively in which the inverse exists when $\bm{X}$ is squared and nonsingular. And the Moore-Penrose generalized inverse of a given $\bm{X}$ with arbitrary structure is denoted as $\bm{X}^{+}$ which has the property that $\left( \bm{X}^{\top} \right)^{+} = \left( \bm{X}^{+} \right)^{\top}$. The notations $\tr(\bm{X})$ and $\diag(\cdots)$ denote the trace of a squared matrix $\bm{X}$ and a diagonal matrix formed by diagonal elements of $\cdots$, respectively. The $n$-dimensional special orthogonal group $SO(n)$ contains all $n$-dimensional rotation matrices and is expressed with ${\bm{X}} \in SO(n) \Rightarrow \bm{X} \in \mathbb{R}^{n \times n}, {\bm{X}}{{\bm{X}}^{\top}} = {{\bm{X}}^{\top}}{\bm{X}} = {\bm{I}},\det ({\bm{X}}) = +1$. The covariance of a given vector $\bm{x}$ perturbed by noises is denoted as $\bm{\Sigma}_{\bm{x}}$. The vectorization of a matrix $\bm{X} = \left(\bm{x}_1, \bm{x}_2, \cdots, \bm{x}_n \right)$ of column size $n$ is defined by $\vecc(\bm{X}) = \left( \bm{x}_1^{\top}, \bm{x}_2^{\top}, \cdots, \bm{x}_n^{\top} \right)^{\top}$ while the inverse operator $\mat[\vecc(\bm{X})]$ restores the vectorization into its original matrix form of $\bm{X}$. The Kronecker product of two arbitrary matrices $\bm{X}$ and $\bm{Y}$ is denoted as $\bm{X} \otimes \bm{Y}$ and commonly $\left( \bm{X} \otimes \bm{Y} \right)^{\top} = \bm{X}^{\top} \otimes \bm{Y}^{\top}$. The convenience of the Kronecker product is that, given matrices $\bm{A}, \bm{B}, \bm{C}, \bm{X}$ with proper sizes, one can write the equality $\bm{AXB} = \bm{C}$ into $\left( \bm{B}^{\top} \otimes \bm{A} \right)\vecc(\bm{X}) = \vecc(\bm{C})$. Also, one may derive $\left( \bm{A} \otimes \bm{B} \right)\left( \bm{C} \otimes \bm{X} \right) = \left( \bm{AC} \otimes \bm{BX} \right)$. For probabilistic descriptions, we use $\left\langle  \cdots  \right\rangle $ to represent the operation of expectation and $\bm{\Sigma}_{\bm{X}}$ or $\bm{\Sigma}_{\bm{XX}}$ is the auto-covariance of a random variable $\bm{X}$. The covariance matrix between two random variables $\bm{X}$ and $\bm{Y}$ are denoted as $\bm{\Sigma}_{\bm{X}, \bm{Y}}$. The notation $\mathcal{N}(\bm{\alpha}, \bm{\Sigma})$ denotes the normal (Gaussian) distribution with mean of $\bm{\alpha}$ and covariance of $\bm{\Sigma}$ while $\mathcal{MN}$ represents the normal distribution for matrix. The probabilistic density function of a given squared matrix $\bm{X}\sim {\mathcal{MN}}\left( \bm{M}, \bm{Y}, \bm{W}\right)$ with dimension $n$ is
\begin{equation}
\begin{gathered}
p\left( \bm{X} | \bm{M}, \bm{Y}, \bm{W}\right) \hfill \\
= \frac{\exp \left\{- \frac{1}{2} \tr \left[ \bm{W}^{-1}(\bm{X} - \bm{M})^{\top} \bm{Y}^{-1}(\bm{X} - \bm{M}) \right]\right\}}{(2\pi )^{\frac{n^2}{2}}\left[\det (\bm{Y}) \det(\bm{W})\right]^{\frac{n}{2}}} \hfill \\
\end{gathered}
\end{equation}
where $\bm{M}$ is the mean of $\bm{X}$ and $\bm{Y}, \bm{W}$ represent the second moments of $\bm{X}$ satisfying
\begin{equation}
\begin{gathered}
\left< (\bm{X} - \bm{M}) (\bm{X} - \bm{M})^{\top}\right> = \bm{Y} \ \tr \ \bm{W} \hfill \\
\left< (\bm{X} - \bm{M})^{\top} (\bm{X} - \bm{M})\right> = \bm{W} \ \tr \ \bm{Y} \hfill \\
\end{gathered}
\end{equation}
The vectorization of $\bm{X}$ will then be subjected to the normal distribution such that
\begin{equation}
\vecc(\bm{X}) \sim {\mathcal{N}}\left[\vecc(\bm{M}), \bm{Y}\otimes\bm{W} \right]
\end{equation}
\hrulefill
%

\begin{figure}[h]
\centering
\includegraphics[width=0.5\textwidth]{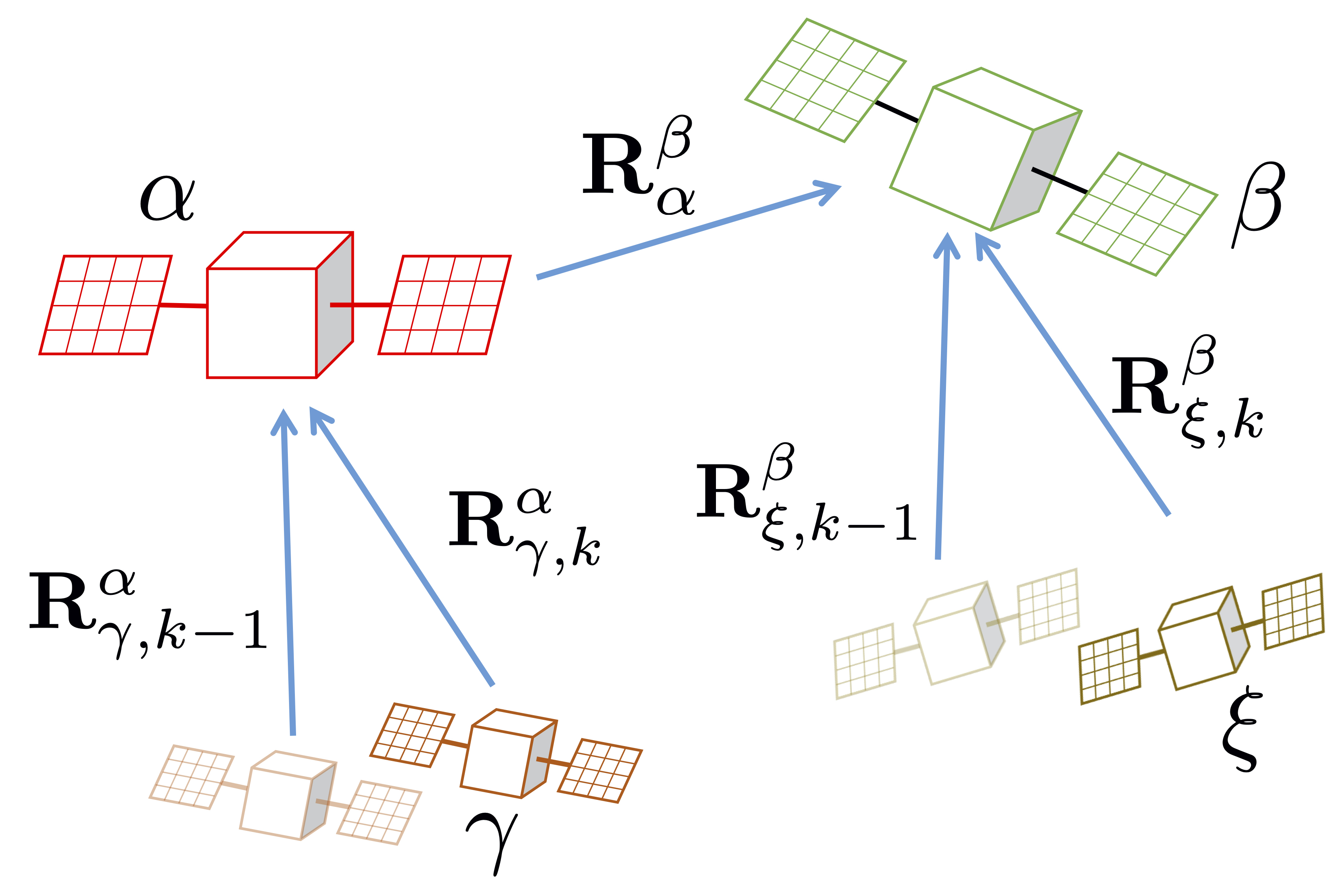}
\caption{Formation-aided relative attitude determination between satellites.}
\label{fig:chart_sats}
\end{figure}

\subsection{The Unified Attitude Determination Problem}
We consider that there are $N$ pairs of $n$-dimensional vector observations and $M$ pairs of $n$-dimensional hand-eye measurements available for attitude determination, all related by the $n$-dimensional rotation matrix $\bm{R} \in SO(n)$: 
\begin{equation}\label{vec_hand}
{\rm{Vector:}}\ \left\{ {\begin{array}{*{20}{c}}
  {{{\bm{b}}_1} = {\bm{R}}{{\bm{r}}_1}} \\ 
  {{{\bm{b}}_2} = {\bm{R}}{{\bm{r}}_2}} \\ 
   \vdots  \\ 
  {{{\bm{b}}_{{N}}} = {\bm{R}}{{\bm{r}}_{{N}}}}
\end{array}} \right., {\rm{Hand-Eye:}}\ \left\{ {\begin{array}{*{20}{c}}
{{{\bm{A}}_1}{\bm{R}} = {\bm{R}}{{\bm{B}}_1}} \\ 
{{{\bm{A}}_2}{\bm{R}} = {\bm{R}}{{\bm{B}}_2}} \\ 
   \vdots  \\ 
  {{{\bm{A}}_{{M}}}{\bm{R}} = {\bm{R}}{{\bm{B}}_{{M}}}} 
  \end{array}} \right.
\end{equation}
where there is no constraint for $\bm{b}_i = \left(b_{i,1}, b_{i,2}, \cdots, b_{i, n} \right)^T, \bm{r}_i = \left(r_{i,1}, r_{i,2}, \cdots, r_{i, n} \right)^T, i = 1, 2, \cdots, N$ and $\bm{A}_i, \bm{B}_i, i = 1, 2, \cdots, M$. That is to say the vector observations do not have to maintain unit norms as presented in the Wahba's problem and hand-eye measurements are also arbitrary and do not have to be symmetric or orthonormal matrices as presented in \cite{Modenini2018} and \cite{Tsai1989}. Fig. \ref{fig:chart_sats} shows a potential example that incorporates the rigid case for $\bm{A}_i$ and $\bm{B}_i$. There are four satellites establishing a formation where the satellite $\alpha$ is conducting attitude synchronization with satellite $\beta$. The aim is to compute the relative attitude $\bm{R}_{\alpha}^{\beta}$ which have already been given by the relative projective geometry using camera with pre-built 3-D models of $\alpha$ and $\beta$. Meanwhile, two passing-through satellites $\gamma$ and $\xi$ also observe the $\alpha$ and $\beta$ respectively. By using the projective geometry likewise, the relative attitude matrices at time instants $k - 1$ and $k$ are estimated. Then using the differential rotation and hand-eye relationship, with the control efforts that $\alpha$ and $\beta$ are attitude-synchronized, we are able to construct
\begin{equation}
\begin{gathered}
\bm{A}_i = \bm{R}_{\gamma, k}^{\alpha} \left( \bm{R}_{\gamma, k - 1}^{\alpha} \right)^\top \hfill \\
\bm{B}_i = \left( \bm{R}_{\xi, k}^{\beta} \right)^\top \bm{R}_{\xi, k - 1}^{\beta} \hfill \\
\end{gathered}
\end{equation}
which denotes the rigid case and will give hand-eye measurements to the original results.\\
\indent Here, the vector observations and hand-eye measurements are assumed to be synchronized that can be satisfied for common data sampling and wireless transmission sub-blocks of the aerospace electronic central system. To derive a general $n$-dimensional closed-form solution to this problem, unlike existing methods used in \cite{Modenini2018}, \cite{Wu2019tim} and \cite{Wu2019sens2}, we use vectorization of rotation matrix for mathematical representation, which has been proven to be effective for hand-eye calibration \cite{Andreff2001,Tabb2017}. For equations of vector observations, we can write them into the compact form
\begin{equation}
\bm{P} = \bm{RQ}
\end{equation}
where 
\begin{equation}
\begin{gathered}
\bm{P} = \left( \bm{b}_1, \bm{b}_2, \cdots, \bm{b}_N\right) \hfill \\
\bm{Q} = \left( \bm{r}_1, \bm{r}_2, \cdots, \bm{r}_N\right) \hfill \\
\end{gathered}
\end{equation}
which can be further vectorized as
\begin{equation}
\vecc(\bm{P}) = \left( \bm{Q}^{\top} \otimes \bm{I} \right) \vecc(\bm{R})
\end{equation}
An analytical solution to the hand-eye calibration problem \cite{Andreff2001} solves the $\bm{R}$ in the hand-eye part of (\ref{vec_hand}) by the following homogeneous system of $\vecc(\bm{R})$
\begin{equation}
\left( 
\begin{gathered}
\bm{I} - \bm{A}_1 \otimes \bm{B}_1 \\
\bm{I} - \bm{A}_2 \otimes \bm{B}_2 \\
\cdots \\
\bm{I} - \bm{A}_M \otimes \bm{B}_M \\
\end{gathered}
\right) \vecc(\bm{R}) = \bm{0}
\end{equation}
This result has been proven to be effective but in fact there is a strong coupling of $\bm{A}_i, \bm{B}_i$ for $i = 1, 2, \cdots, M$ which sets obstacle for a further covariance analysis. Therefore, we use another intuitive solution such that
\begin{equation}
\left( 
\begin{gathered}
\bm{A}_1 \otimes \bm{I} - \bm{I} \otimes \bm{B}_1^{\top} \\
\bm{A}_2 \otimes \bm{I} - \bm{I} \otimes \bm{B}_2^{\top} \\
\cdots \\
\bm{A}_M \otimes \bm{I} - \bm{I} \otimes \bm{B}_M^{\top} \\
\end{gathered}
\right) \vecc(\bm{R}) = \bm{0}
\end{equation}
The unified attitude determination result from vector observations and hand-eye measurements then can be given by the following problem
\begin{equation}\label{con_problem}
\begin{gathered}
\mathop{\arg \min} \limits_{\bm{x}} \ {\mathcal{L}} = \bm{x}^{\top}\bm{H} \bm{x} + \hfill \\
\ \ \ \ \ \ \ \ \ \ \ \ \left[\left(\bm{Q}^{\top} \otimes \bm{I}\right) \bm{x} - \vecc(\bm{P}) \right]^{\top}\left[\left(\bm{Q}^{\top} \otimes \bm{I}\right) \bm{x} - \vecc(\bm{P}) \right] \hfill \\
\end{gathered}
\end{equation}
in which $\bm{x} = \vecc(\bm{R})$ and 
\begin{equation}
\bm{H} = \sum \limits_{i = 1}^{M} \left(\bm{A}_i \otimes \bm{I} - \bm{I} \otimes \bm{B}_i^{\top} \right)^{\top}\left(\bm{A}_i \otimes \bm{I} - \bm{I} \otimes \bm{B}_i^{\top} \right)
\end{equation}
It should be noted that if the weights of the measurements are taken into account, we should use the following matrices instead:
\begin{equation}
\begin{gathered}
\bm{P} = \left( \sqrt{w_1}\bm{b}_1, \sqrt{w_2}\bm{b}_2, \cdots, \sqrt{w_N}\bm{b}_N\right) \hfill \\
\bm{Q} = \left( \sqrt{w_1}\bm{r}_1, \sqrt{w_2}\bm{r}_2, \cdots, \sqrt{w_N}\bm{r}_N\right) \hfill \\
\bm{H} = \sum \limits_{i = 1}^{M} v_i\left(\bm{A}_i \otimes \bm{I} - \bm{I} \otimes \bm{B}_i^{\top} \right)^{\top}\left(\bm{A}_i \otimes \bm{I} - \bm{I} \otimes \bm{B}_i^{\top} \right) \hfill \\
\end{gathered}
\end{equation}
with $w_1, w_2, \cdots, w_N$ the positive weights of $N$ vector measurements and $v_1, v_2, \cdots, v_M$ the positive weights of $M$ hand-eye measurements. The ratio $\varrho = \left(\sum\limits_{i = 1}^{N}w_i \right) / \left ( \sum\limits_{i = 1}^{M}v_i \right )$ describes the relative accuracy between the vector and hand-eye measurements. The objective function $\mathcal{L}$ can be evaluated as
\begin{equation}
\begin{gathered}
\ {\mathcal{L}} = \bm{x}^{\top}\left[ \bm{H}  + \left(\bm{Q} \otimes \bm{I}\right)\left(\bm{Q}^{\top} \otimes \bm{I}\right) \right]\bm{x} - \hfil \\
\ \ \ \ \ \ \ \ \ \ \ \bm{x}^{\top} \left(\bm{Q} \otimes \bm{I}\right)^{\top}\vecc(\bm{P}) - \hfill \\ 
\ \ \ \ \ \ \ \ \ \ \ \vecc(\bm{P})^{\top} \left(\bm{Q}^{\top} \otimes \bm{I}\right)\bm{x} + \vecc(\bm{P})^{\top} \vecc(\bm{P}) \hfill \\
\end{gathered}
\end{equation}
The optimal $\bm{x}$ occurs at
\begin{equation} \label{partial}
\begin{gathered}
\frac{\partial {\mathcal{L}}}{\partial \bm{x}} = 2 \bm{x}^{\top}\left[ \bm{H}  + \left(\bm{Q} \otimes \bm{I}\right)\left(\bm{Q}^{\top} \otimes \bm{I}\right) \right] - \hfill \\
\ \ \ \ \ \ \ \ \ 2\vecc(\bm{P})^{\top}\left(\bm{Q}^{\top} \otimes \bm{I}\right) \hfill \\
\ \ \ \ \ = \bm{0} \hfill \\
\end{gathered}
\end{equation}
which generates
\begin{equation}\label{x_sol}
\bm{x} = \left[ \bm{H}  + \left(\bm{Q} \otimes \bm{I}\right)\left(\bm{Q}^{\top} \otimes \bm{I}\right) \right]^{+} \left(\bm{Q} \otimes \bm{I}\right)\vecc(\bm{P})
\end{equation}
The obtained result indicates that the general observability of attitude angles are governed by the rank of $\left[ \bm{H}  + \left(\bm{Q} \otimes \bm{I}\right)\left(\bm{Q}^{\top} \otimes \bm{I}\right) \right]$. When there is no vector observation, (\ref{x_sol}) can not hold since $\left(\bm{Q} \otimes \bm{I}\right)\vecc(\bm{P})$ becomes null. However, when there is no hand-eye measurement, (\ref{x_sol}) could also make sense which depends on the number of vector observations and the collinearity between vector pairs \cite{Wu2017iet,Wu2018iet,Markley2018}. That is so say, in this way the following equation is also a Wahba's solution
\begin{equation}
\bm{x} = \left[\left(\bm{Q} \otimes \bm{I}\right)\left(\bm{Q}^{\top} \otimes \bm{I}\right) \right]^{+} \left(\bm{Q} \otimes \bm{I}\right)\vecc(\bm{P})
\end{equation}
For instance, with only two vector observation pairs, we may have already compute the attitude \cite{Markley2002}. When there are one vector observation pair and one set of hand-eye measurements, there could be sufficient information for a full-attitude determination since the single system $\bm{AR} = \bm{RB}$ has been proven to own two ambiguous solutions \cite{Modenini2018} and the optimal can be then selected by integrating an external vector pair.\\
\indent When there are only hand-eye measurements, the optimization (\ref{con_problem}) will be
\begin{equation}\label{hand_eye_only}
\mathop{\arg \min} \limits_{\bm{x}} \ {\mathcal{L}} = \bm{x}^{\top}\bm{H} \bm{x}
\end{equation} 
indicating that $\bm{x}$ is the eigenvector of $\bm{H}$ associated with its minimum eigenvalue. As $\bm{x}$ and $ - \bm{x}$ are all such eigenvectors, it is able to verify the reconstructed $\bm{R} = \mat (\bm{x})$ via loss function $\mathcal{L}$. \\
\indent The obtained attitude reconstruction, since may be put into a further Kalman filter for more accurate state estimation and gyro bias cancellation, should provide its uncertainty descriptions \cite{Zhou2013,Zhou2019}. The detailed derivations are presented in the next sub-section.

\subsection{Covariance Analysis}
In this sub-section, the covariance analysis of the derived solution (\ref{x_sol}) will be performed. Note that when there are only hand-eye measurements, the problem degenerates to (\ref{hand_eye_only}). In such case, the uncertainty descriptions of $\bm{x}$ are available via \cite{Liounis2016}. From (\ref{partial}), one easily observes that the solution of $\bm{x}$ is linear. Thus a perturbed model of $\bm{x}$ can be obtained by
\begin{equation}
\begin{small}
\begin{gathered}
\left[ \bm{H}  + \left(\bm{Q} \otimes \bm{I}\right)\left(\bm{Q}^{\top} \otimes \bm{I}\right) \right] \delta \bm{x} + 
\delta \left[ \bm{H}  + \left(\bm{Q} \otimes \bm{I}\right)\left(\bm{Q}^{\top} \otimes \bm{I}\right) \right] \bm{x} \hfill \\
\ \ \ \ \ = \delta \left[ \left(\bm{Q} \otimes \bm{I}\right)\vecc(\bm{P})\right] \hfill \\
\Rightarrow \left[ \bm{H}  + \left(\bm{Q} \otimes \bm{I}\right)\left(\bm{Q}^{\top} \otimes \bm{I}\right) \right] \delta \bm{x} + \hfill \\
\left[
\begin{gathered}
\delta \bm{H}  + \left(\delta\bm{Q} \otimes \bm{I}\right)\left(\bm{Q}^{\top} \otimes \bm{I}\right) + \\
\left(\bm{Q} \otimes \bm{I}\right)\left(\delta\bm{Q}^{\top} \otimes \bm{I}\right) 
\end{gathered}
\right] \bm{x} \\
\ \ \ \ \ = \left(\delta\bm{Q} \otimes \bm{I}\right)\vecc(\bm{P}) + \left(\bm{Q} \otimes \bm{I}\right)\vecc(\delta \bm{P}) \hfill \\
\end{gathered}
\end{small}
\end{equation}
where the second-order differential terms are ommited. The covariance of $\bm{x}$ can be computed via
\begin{equation}
\begin{small}
\begin{gathered}
\bm{\Sigma}_{\bm{x}\bm{x}} = \left[ \bm{H}  + \left(\bm{Q} \otimes \bm{I}\right)\left(\bm{Q}^{\top} \otimes \bm{I}\right) \right]^{+} \times \hfill \\
\ \ \ \ \ \left\langle
\begin{gathered}
\left\{ 
\begin{gathered}
\left[ \left(\delta\bm{Q} \otimes \bm{I}\right)\vecc(\bm{P}) + \left(\bm{Q} \otimes \bm{I}\right)\vecc(\delta \bm{P})\right] - \hfill \\ 
\left[
\begin{gathered}
\delta \bm{H}  + \left(\delta\bm{Q} \otimes \bm{I}\right)\left(\bm{Q}^{\top} \otimes \bm{I}\right) + \hfill \\
\left(\bm{Q} \otimes \bm{I}\right)\left(\delta\bm{Q}^{\top} \otimes \bm{I}\right) \hfill \\
\end{gathered}
\right] \bm{x} \hfill \\
\end{gathered}
\right\} \\
\left\{ 
\begin{gathered}
\left[ \left(\delta\bm{Q} \otimes \bm{I}\right)\vecc(\bm{P}) + \left(\bm{Q} \otimes \bm{I}\right)\vecc(\delta \bm{P})\right] - \hfill \\ 
\left[
\begin{gathered}
\delta \bm{H}  + \left(\delta\bm{Q} \otimes \bm{I}\right)\left(\bm{Q}^{\top} \otimes \bm{I}\right) + \hfill \\
 \left(\bm{Q} \otimes \bm{I}\right)\left(\delta\bm{Q}^{\top} \otimes \bm{I}\right) \hfill \\
\end{gathered}
\right] \bm{x} \hfill \\
\end{gathered}
\right\}^{\top}
\end{gathered}
\right\rangle \hfill \\
\ \ \ \ \ \times \left[ \bm{H}  + \left(\bm{Q} \otimes \bm{I}\right)\left(\bm{Q}^{\top} \otimes \bm{I}\right) \right]^{+} \hfill \\
\end{gathered}
\end{small}
\end{equation}
As the operation of the type $\left(\delta\bm{Q} \otimes \bm{I}\right)\vecc(\bm{P})$ is linear both in elements of $\bm{Q}$ and $\bm{P}$, one can write the matrix manipulations into
\begin{equation}
\begin{gathered}
\left(\delta\bm{Q} \otimes \bm{I}\right)\vecc(\bm{P}) = {\mathcal{Z}}(\bm{P}) \vecc(\delta \bm{Q}^{\top}) 
\end{gathered}
\end{equation}
where ${\mathcal{Z}}(\bm{P})$ a linearly mapping. Note that
\begin{equation}
\begin{gathered}
\mat\left[\left(\delta\bm{Q} \otimes \bm{I}\right)\vecc(\bm{P}) \right] = \bm{P} \delta \bm{Q}^{\top} \hfill \\
\Rightarrow \vecc\left( \bm{P} \delta \bm{Q}^{\top}\right) = \left( \bm{I} \otimes \bm{P} \right)\vecc\left( \delta \bm{Q}^{\top}\right)
\end{gathered}
\end{equation}
then we have
\begin{equation}
{\mathcal{Z}} \left( \bm{P} \right) = \left( \bm{I} \otimes \bm{P} \right)
\end{equation}
In this way, one has

\begin{figure*}[hb]
\centering
\includegraphics[width=0.65\textwidth]{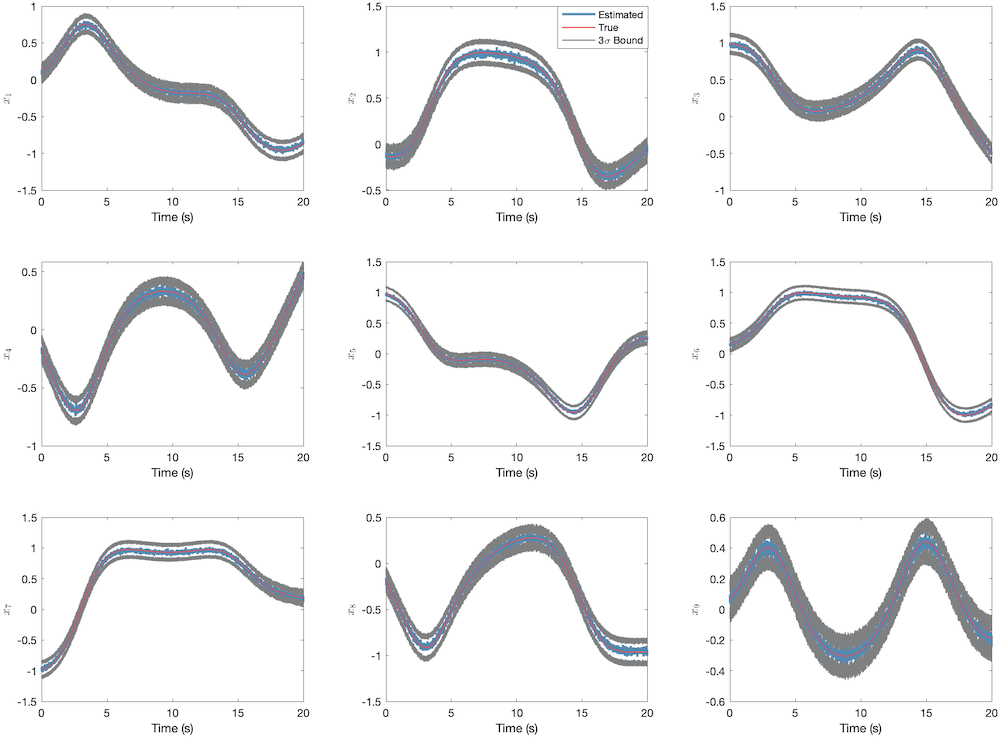}
\caption{Case 1: The estimated results using symmetric $\bm{A}_i, \bm{B}_i$ and their $3\sigma$ bounds when $\epsilon_{\rm{vector}} = 1 \times 10^{-1}$, $\epsilon_{\rm{hand-eye}} = 1 \times 10^{-5}$, $N = 30$ and $M = 1$.}
\label{fig:cov2}
%
\centering
\includegraphics[width=0.65\textwidth]{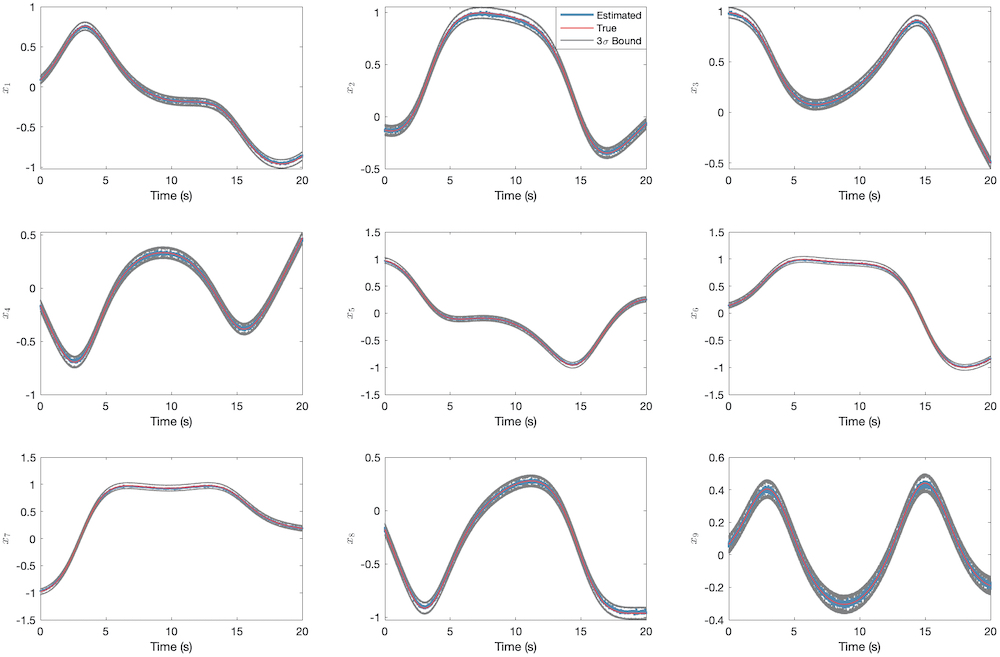}
\caption{Case 1: The estimated results using symmetric $\bm{A}_i, \bm{B}_i$ and their $3\sigma$ bounds when $\epsilon_{\rm{vector}} = 1 \times 10^{-1}$, $\epsilon_{\rm{hand-eye}} = 1 \times 10^{-5}$, $N = 150$ and $M = 1$.}
\label{fig:cov1}
\end{figure*}

\begin{equation}
\begin{gathered}
\left(\delta\bm{Q} \otimes \bm{I}\right)\left(\bm{Q}^{\top} \otimes \bm{I}\right)\bm{x} \hfill \\
= {\mathcal{Z}}\left\{ \mat\left[\left(\bm{Q}^{\top} \otimes \bm{I}\right)\bm{x}\right]\right\}] \vecc(\delta \bm{Q}^{\top}) \hfill \\
= {\mathcal{Z}}\left( \bm{\tilde RQ} \right) \vecc(\delta \bm{Q}^{\top}) \hfill \\
\end{gathered}
\end{equation}
where $\bm{\tilde R} = \mat \left[ \vecc(\bm{x})\right]$ is the reconstructed rotation matrix and is not strictly on the $SO(n)$. In the same manner, we can also obtain
\begin{equation}
\begin{gathered}
\delta \bm{Hx} \hfill \\
= \sum \limits_{i = 1}^{M} v_i\left[
\begin{gathered}
\left(\delta \bm{A}_i \otimes \bm{I} - \bm{I} \otimes \delta \bm{B}_i^{\top} \right)^{\top}\left(\bm{A}_i \otimes \bm{I} - \bm{I} \otimes \bm{B}_i^{\top} \right) + \\
\left(\bm{A}_i \otimes \bm{I} - \bm{I} \otimes \bm{B}_i^{\top} \right)^{\top}\left(\delta \bm{A}_i \otimes \bm{I} - \bm{I} \otimes \delta \bm{B}_i^{\top} \right) 
\end{gathered}
\right]\bm{x} \hfill \\
= \sum \limits_{i = 1}^{M}  v_i \left\{ 
\begin{gathered}
{\mathcal{F}}\left[\left(\bm{A}_i \otimes \bm{I} - \bm{I} \otimes \bm{B}_i^{\top} \right)\bm{x}\right] 
\left[
\begin{gathered}
\vecc(\delta \bm{A}_i)\\
\vecc(\delta \bm{B}_i)\\
\end{gathered} 
\right] + \\
 \left(\bm{A}_i^{\top} \otimes \bm{I} - \bm{I} \otimes \bm{B}_i \right){\mathcal{F}}(\bm{x})
\left[
\begin{gathered}
\vecc(\delta \bm{A}_i^{\top})\\
\vecc(\delta \bm{B}_i^{\top})\\
\end{gathered} 
\right] 
\end{gathered}
\right\} \hfill \\
\end{gathered}
\end{equation}
where the function $\mathcal{F}(\bm{x})$ is a linear mapping which can be derived as follows
\begin{equation}
\begin{gathered}
\left(\delta \bm{A} \otimes \bm{I} - \bm{I} \otimes \delta \bm{B}^{\top} \right)^{\top}\vecc({\bm{C}}) \hfill \\
= \left(\delta \bm{A}^{\top} \otimes \bm{I} - \bm{I} \otimes \delta \bm{B} \right)\vecc({\bm{C}}) \hfill \\
= \left( \delta \bm{A}^{\top} \otimes \bm{I} \right) \vecc \left( \bm{C}\right) - \left( \bm{I} \otimes \delta \bm{B} \right) \vecc({\bm{C}}) \hfill \\
= \vecc \left( \bm{C} \delta \bm{A} \right) - \vecc \left( \delta \bm{B} \bm{C} \right) \hfill \\
= \left( \bm{I} \otimes \bm{C} \right) \vecc(\delta \bm{A}) - \left( \bm{C}^{\top} \otimes \bm{I} \right) \vecc(\delta \bm{B}) \hfill \\
= \left(
\begin{gathered}
\bm{I} \otimes \bm{C}, 
- \bm{C}^{\top} \otimes \bm{I}
\end{gathered}
\right) \left[
\begin{gathered}
\vecc(\delta \bm{A})\\
\vecc(\delta \bm{B})\\
\end{gathered} 
\right ] \hfill \\
\Rightarrow {\mathcal{F}} \left [ \vecc(\bm{C})\right] = \left(
\begin{gathered}
\bm{I} \otimes \bm{C}, 
- \bm{C}^{\top} \otimes \bm{I}
\end{gathered}
\right) \hfill \\
\end{gathered}
\end{equation}

\begin{figure*}[ht]
\centering
\includegraphics[width=0.65\textwidth]{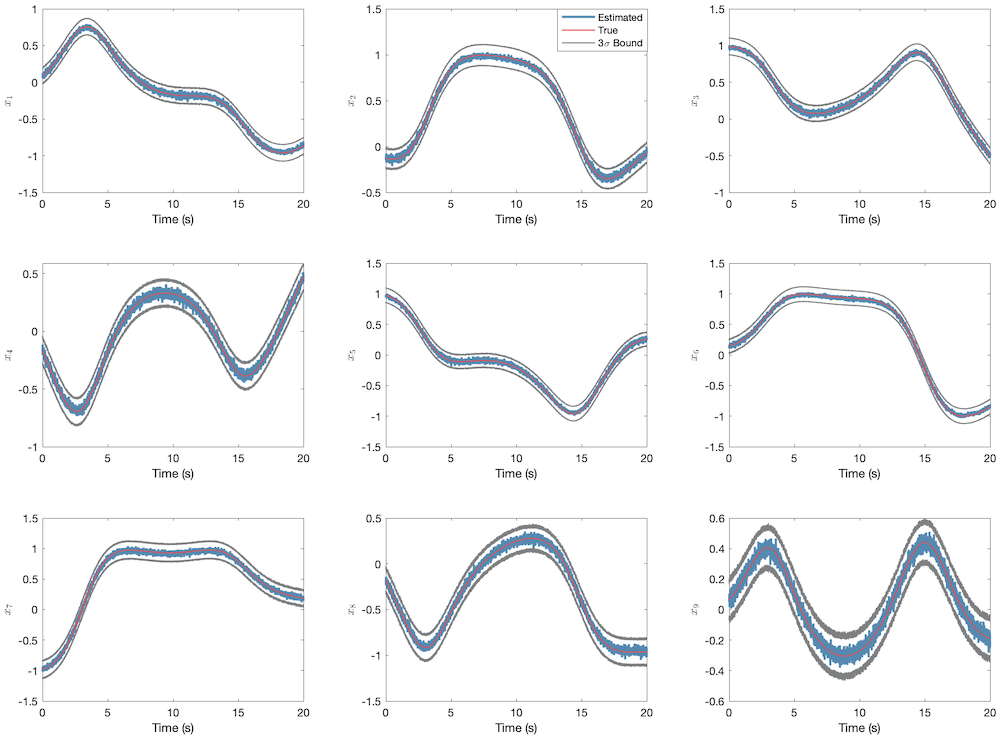}
\caption{Case 2: The estimated results using rigid $\bm{A}_i, \bm{B}_i$ and their $3\sigma$ bounds when $\epsilon_{\rm{vector}} = 1 \times 10^{-1}$, $\epsilon_{\rm{hand-eye}} = 1 \times 10^{-5}$, $N = 30$ and $M = 1$.}
\label{fig:cov4}
%
\centering
\includegraphics[width=0.65\textwidth]{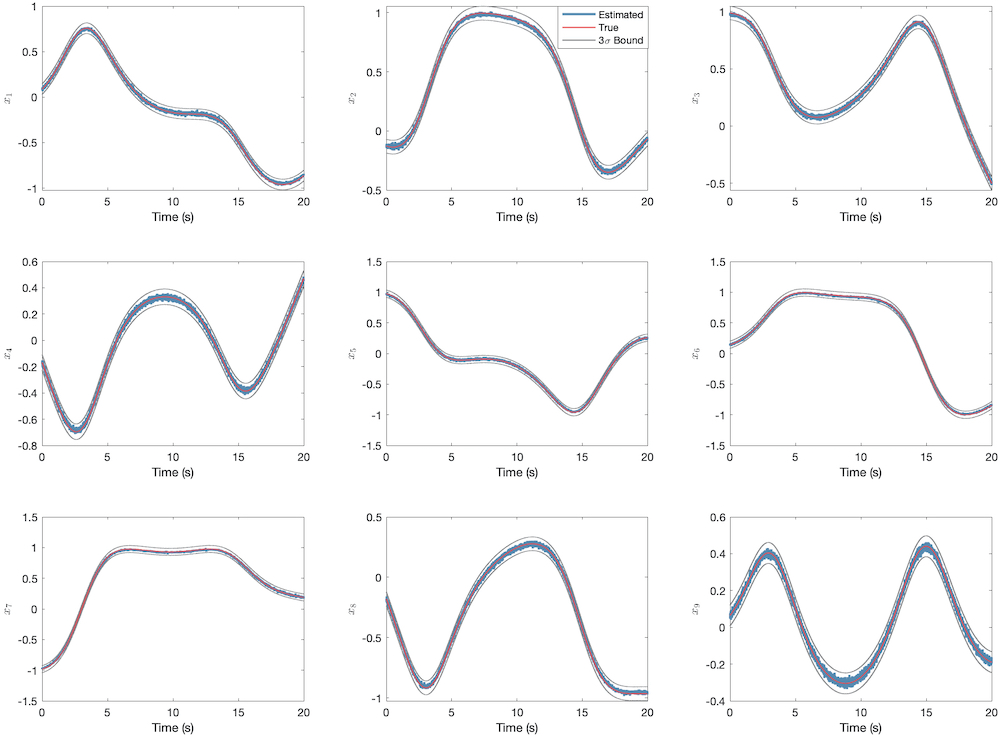}
\caption{Case 2: The estimated results using rigid $\bm{A}_i, \bm{B}_i$ and their $3\sigma$ bounds when $\epsilon_{\rm{vector}} = 1 \times 10^{-1}$, $\epsilon_{\rm{hand-eye}} = 1 \times 10^{-5}$, $N = 150$ and $M = 1$.}
\label{fig:cov3}
\end{figure*}

Thus, the simplified form of $\bm{\Sigma}_{\bm{x}\bm{x}}$ is presented by
\begin{equation}\label{cov_x}
\begin{gathered}
\bm{\Sigma}_{\bm{xx}} =  \left[ \bm{H}  + \left(\bm{Q} \otimes \bm{I}\right)\left(\bm{Q}^{\top} \otimes \bm{I}\right) \right]^{+} \times \hfill \\
\ \ \ \ \ \ \ \ \ \ \left( \bm{S}_1 + \bm{S}_2 + \bm{S}_3 + \bm{S}_3^{\top} \right) \left[ \bm{H}  + \left(\bm{Q} \otimes \bm{I}\right)\left(\bm{Q}^{\top} \otimes \bm{I}\right) \right]^{+} \hfill \\
\end{gathered}
\end{equation}
\noindent in which the detailed derivations along with $\bm{S}_1, \bm{S}_2, \bm{S}_3$ are in the appendix. Here, we need to note that $\bm{x} = \vecc(\bm{R})$ is the vectorization of $\bm{R}$ and to get a proper $\bm{R}$ in $SO(n)$, one should orthonormalize $\mat\left(\bm{x}\right)$ from (\ref{x_sol}). Typically, such normalization is achieved by
\begin{equation}
\bm{R} = \bm{U}\diag\left[1, 1, \cdots, \det(\bm{UV})\right] \bm{V}^{\top}
\end{equation}
where $\bm{USV}^{\top} = \mat\left(\bm{x}\right)$ is the singular value decomposition (SVD) of $\mat\left(\bm{x}\right)$. When $\bm{R} \in SO(3)$, the orthonormalization can also be conducted by solutions to Wahba's problem \cite{Bar-Itzhack2000} that involves an eigenvalue problem of $4 \times 4$ matrices. As SVD is highly nonlinear, we introduce a recent linear method for generalized $n$-dimensional registration \cite{Wu2019glnr} with uncertainty descriptions so that the following system can be established
\begin{equation}\label{glnr}
\left\{ {\begin{array}{*{20}{c}}
  {{{\bm{d}}_1} = {\bm{R}}{{\bm{e}}_1}} \\ 
  {{{\bm{d}}_2} = {\bm{R}}{{\bm{e}}_2}} \\
   \vdots  \\ 
  {{{\bm{d}}_{{n}}} = {\bm{R}}{{\bm{e}}_{{n}}}}
\end{array}} \right.
\end{equation}
where $\bm{x} = \vecc(\bm{R}) = \left( \bm{d}_1^{\top}, \bm{d}_2^{\top}, \cdots, \bm{d}_n^{\top}\right)^{\top}$ and $\bm{e}_1 = \left(1, 0, 0, \cdots, 0 \right)^{\top}$, $\bm{e}_2 = \left(0, 1, 0, \cdots, 0 \right)^{\top}$, $\cdots$, $\bm{e}_n = \left(0, \cdots,0, 1 \right)^{\top}$ are standard orthogonal unit bases of Euclidean space $\mathbb{R}^n$. In such model, the vectors $\bm{e}_1, \bm{e}_2, \cdots, \bm{e}_n$ are noise-free and the covariance of $\bm{d}_1, \bm{d}_2, \cdots, \bm{d}_n$ are invoked directly from $\bm{\Sigma}_{\bm{xx}}$, so that the covariance $\bm{\Sigma}_{\bm{R}}$ is obtained. (\ref{glnr}) frames a standard form of point-cloud registration, and $\bm{R}$ is its optimal solution on $SO(n)$. First, we need to reconstruct the matrix
\begin{equation}
\bm{G} = \sum \limits_{i = 1}^n \frac{1}{n} {\mathcal{P}}^\top (\bm{\tau}_i){\mathcal{P}} (\bm{\tau}_i)
\end{equation}
where $\bm{\tau}_i = \bm{d}_i + \bm{e}_i$. Then reconstruct the vector
\begin{equation}
\bm{v} = \sum \limits_{i = 1}^n \frac{1}{n} {\mathcal{P}}^\top (\bm{\tau}_i) \bm{\varrho}_i
\end{equation}
where $\bm{\varrho} = \bm{e}_i - \bm{d}_i$. $\bm{R}$ satisfies the following Caylay transform
\begin{equation}
\bm{R} = \left( \bm{I} + \bm{g}_{\odot} \right)^{-1} \left( \bm{I} - \bm{g}_{\odot} \right)
\end{equation}
where $\bm{g}_{\odot}$ gives a skew-symmetric matrix by a linear mapping from ${\bm{g}} = {\left[ {{g_1},{g_2}, \cdots ,{g_{\frac{{n(n - 1)}}{2}}}} \right]^{\top}}$, such that
\begin{equation}
\begin{gathered}
\bm{g}_{\odot} = \hfill \\
\left( {\begin{array}{*{20}{c}}
  0&{{g_1}}&{{g_2}}& \cdots &{{g_{n - 1}}} \\ 
  { - {g_1}}&0&{{g_n}}& \cdots &{{g_{2n - 3}}} \\ 
  { - {g_2}}&{{g_n}}& \ddots & \cdots & \vdots  \\ 
   \vdots & \vdots & \vdots &0&{{g_{\frac{{n\left( {n - 1} \right)}}{2}}}} \\ 
  { - {g_{n - 1}}}&{ - {g_{2n - 3}}}& \cdots &{ - {g_{\frac{{n\left( {n - 1} \right)}}{2}}}}&0 
\end{array}} \right) \hfill \\
\end{gathered}
\end{equation}
The matrix $\mathcal{P}$ meets the following transform 
\begin{equation}
\bm{g}_{\odot} \bm{\tau}_i = {\mathcal{P}} (\bm{\tau}_i) \bm{g}
\end{equation}
and can be evaluated via various symbolic engines like MATLAB and Mathematica. Then the covariance of $\bm{g}$ i.e. $\bm{\Sigma}_{\bm{g}}$ can be given by (43) in \cite{Wu2019glnr}. After that, the rotation uncertainty can be characterized by means of
\begin{equation}
\bm{\Sigma}_{\bm{R}} = (\bm{I} + \bm{g}_{\odot})^{-1} \left[ {\sum\limits_{i = 1}^n {{\mathcal{P}}\left( {{{\bm{\zeta}}_i}} \right){{\bm{\Sigma}}_{\bm{g}}}{{\mathcal{P}}^{\top}}\left( {{{\bm{\zeta}}_i}} \right)} } \right]{\left[ {{{\left( {{\bm{I}} + \bm{g}_{\odot}} \right)}^{ - 1}}} \right]^{\top}}
\end{equation}
in which
\begin{equation}
{\bm{R}} + {\bm{I}} = \left( {{{\bm{\zeta}}_1},{{\bm{\zeta}}_2}, \cdots ,{{\bm{\zeta}}_n}} \right)
\end{equation}

\section{Experimental Results}
\subsection{Simulation Results}
\begin{figure}[h]
\centering
\includegraphics[width=0.5\textwidth]{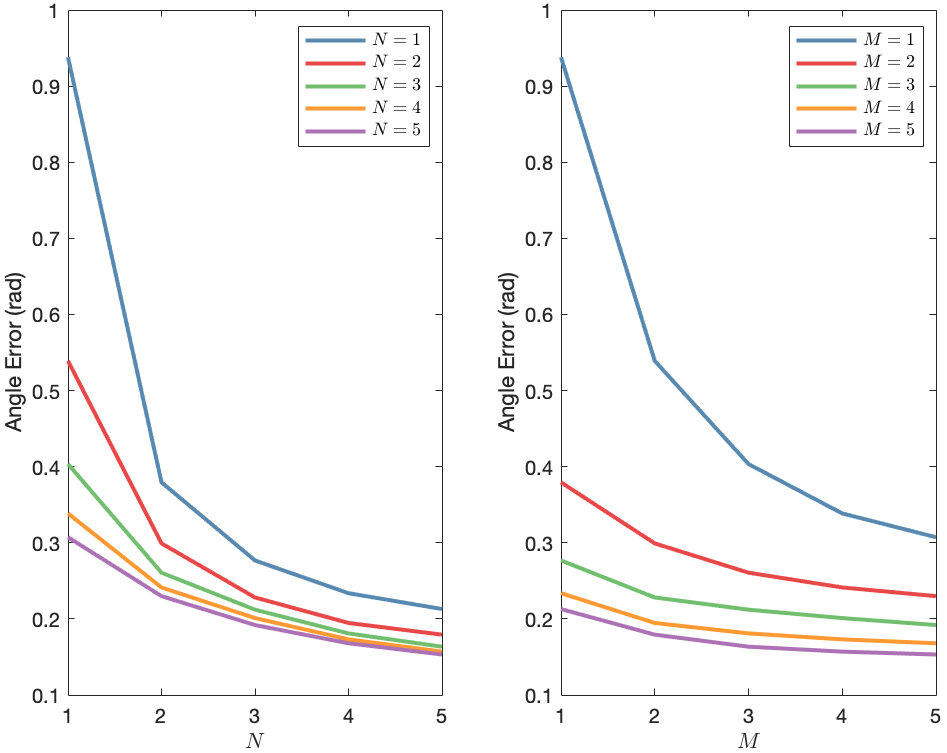}
\caption{The rotation errors subject to different values of $N$ and $M$ with $\bm{A}_i, \bm{B}_i \in SO(n), n = 3$ (rigid case).}
\label{fig:accuracy}
\end{figure}

\begin{figure}[h]
\centering
\includegraphics[width=0.5\textwidth]{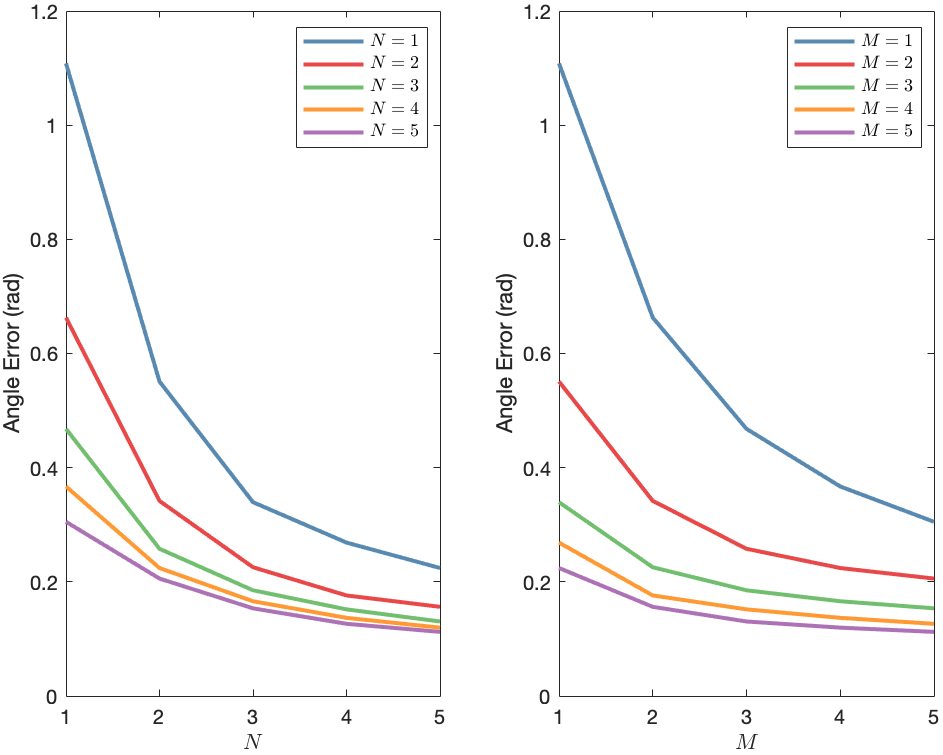}
\caption{The rotation errors subject to different values of $N$ and $M$ with $\bm{A}_i, \bm{B}_i$ being $3 \times 3$ symmetric matrices.}
\label{fig:accuracy_sym}
\end{figure}

In this sub-section, we simulate several cases to demonstrate the performances of the proposed algorithm. The rotation error is defined as
\begin{equation}
\eta = {\rm{arc}} \cos \frac{\tr (\bm{R} \bm{R}_{\true}^\top) - 1}{2}
\end{equation}
with $\bm{R}_{\true}$ the reference (true) rotation matrix and $\epsilon_{\rm{vector}}, \epsilon_{\rm{hand-eye}}$ denoting the noise densities of the vector observations and hand-eye measurements respectively. The vector observations and hand-eye measurements are simulated using additive noises such that
\begin{equation}
\begin{gathered}
\bm{b}_i = \bm{R}_{\true} \bm{r}_i + \bm{\varepsilon}_{i} \hfill \\
\bm{A}_i \bm{R}_{\true} - \bm{R}_{\true} \bm{B}_i = \bm{\Xi}_{i} \hfill \\
\end{gathered}
\end{equation}
with $\bm{\varepsilon}_i$ and $\bm{\Xi}_i$ the noises subject to Gaussian distribution such that
\begin{equation}
\begin{gathered}
\bm{\varepsilon}_i \sim {\mathcal{N}}\left(\bm{0},  \epsilon_{\rm{vector}} \bm{I}\right) \hfill \\
\bm{\Xi}_i \sim {\mathcal{MN}}\left(\bm{0},  \epsilon_{\rm{hand-eye}} \bm{I}, \epsilon_{\rm{hand-eye}} \bm{I}\right) \hfill \\
\end{gathered}
\end{equation}
The sequence of the true rotation matices is generated using a modified quaternion model in \cite{Wu2018jgcd}:
\begin{equation}
\begin{gathered}
{\bm{q}} = \left[ {\begin{array}{*{20}{c}}
{\sin \left( { - 0.8334\times 2 \times{10^{ - 3}}k + {\rm{1.3679}}} \right)}\\
{\sin \left( { - 1.5833\times 2 \times{10^{ - 3}}k - 0.1479} \right)}\\
{\sin \left( {3.0038\times 2 \times{10^{ - 3}}k + 2.0061} \right)}\\
{\sin \left( { - 1.1200\times 2 \times{10^{ - 3}}k - 0.0179} \right)}
\end{array}} \right] , {\bm{q}} = \frac{\bm{q}}{\sqrt{\bm{q}^{\top}\bm{q}}}
\end{gathered}
\end{equation}
where $k = 1, 2, \cdots, 10000$ denote the indices. We validate our algorithm by converting the quaternions into rotation matrices. The vector observations are assumed to have been obtained by rotation matching from feature extraction algorithms. Since the feature extraction may be opportunistic, the single-point covariance parameter of the vector observations is set to $\epsilon_{\rm{vector}} = 1 \times 10^{-1}$. While in \cite{Modenini2019}, it is reported that the hand-eye measurements can reach the attitude determination accuracy up to several arc seconds, $\epsilon_{\rm{hand-eye}}$ is set to $1 \times 10^{-5}$ i.e. this hand-eye source is regarded much more accurate than the feature-matching based approach. It is assumed that here only one asteroid is in the field-of-view (FOV) of the camera so $M = 1$. For the first set of cases, we simulate the symmetric $\bm{A}_i, \bm{B}_i$ as appeared in \cite{Modenini2018,Modenini2019}. Two cases with different vector observation numbers $N = 30$ and $N = 150$ are simulated. The results along with the $3\sigma$ bounds calculated from the covariance matrices are shown in Fig. \ref{fig:cov2} and \ref{fig:cov1}. For another set of cases, $\bm{A}_i, \bm{B}_i$ are rigid ones like those in the hand-eye calibration while the results are presented in Fig. \ref{fig:cov4} and \ref{fig:cov3}. \\
\indent From the two sets of cases, one can observe that with fewer vector observations, the attitude estimator has worse accuracy and the uncertainty of the $3\sigma$ bounds will be larger. These $3\sigma$ bounds are taken from the square roots of the diagonal elements of the covariance matrix. Comparing the first and second sets of cases, we can also see that the system with symmetric $\bm{A}_i, \bm{B}_i$ owns worse accuracy than those with rigid hand-eye measurements. To study the relationship between the rotation error and numbers of vector observations and hand-eye measurements, we choose $N$ and $M$ from 1 to 5 for simulation. In each step, the vector observations and hand-eye measurements are randomly sampled for 5000 times and the results are averaged so the statistical trends can be well reflected. The noises are chosen to be subjected to
\begin{equation}
\begin{gathered}
\bm{\varepsilon}_i \sim {\mathcal{N}}\left(\bm{0},  5\times 10^{-1} \bm{I}\right) \hfill \\
\bm{\Xi}_i \sim {\mathcal{MN}}\left(\bm{0},  5\times 10^{-1} \bm{I}, 5\times 10^{-1} \bm{I}\right) \hfill \\
\end{gathered}
\end{equation}
i.e. both types of measurements are very noisy so the error scales can be magnified. We also study two cases where $\bm{A}_i, \bm{B}_i$ are rigid or symmetric and the relationships are then shown in Fig. \ref{fig:accuracy} and \ref{fig:accuracy_sym}.

\begin{figure*}[htbp]
\centering
\includegraphics[width=0.8\textwidth]{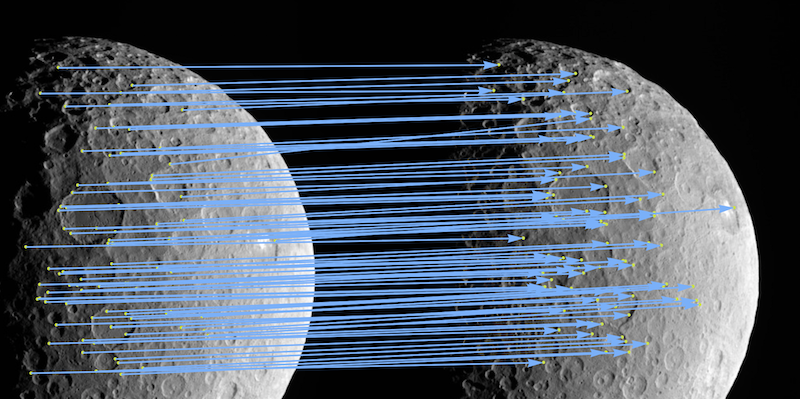}
\caption{The matched features in successive images of Ceres from the Dawn spacecraft using BRISK.}
\label{fig:brisk}
\centering
\includegraphics[width=0.8\textwidth]{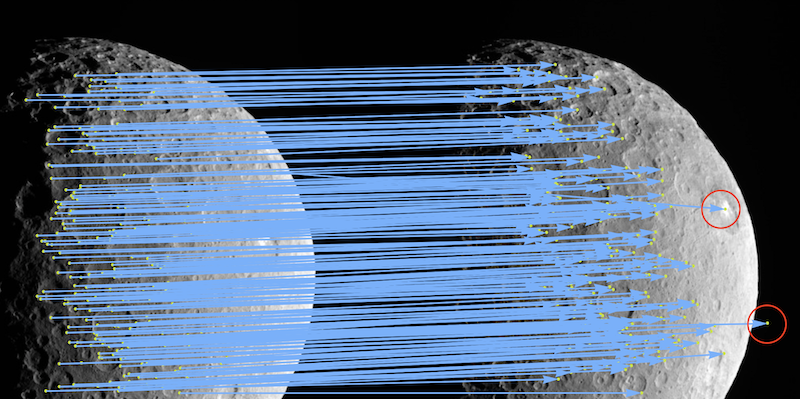}
\caption{The matched features in successive images of Ceres from the Dawn spacecraft using SURF.}
\label{fig:surf}
\centering
\includegraphics[width=0.8\textwidth]{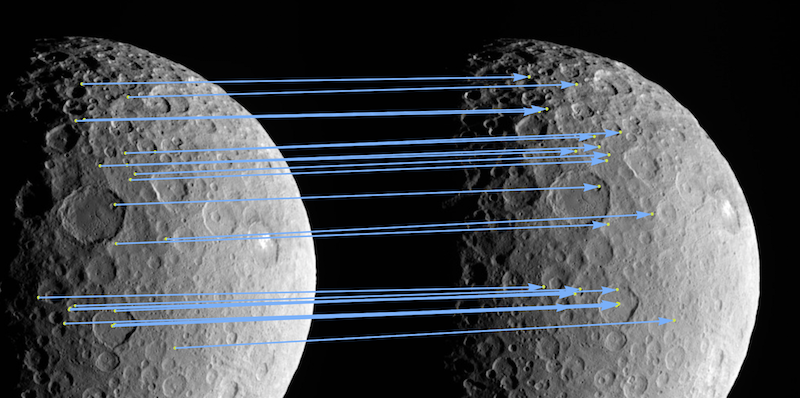}
\caption{The matched features in successive images of Ceres from the Dawn spacecraft using ORB.}
\label{fig:orb}
\end{figure*}

\indent With growing $N$ and $M$, the rotation errors rapidly decrease while the decreasing ratio for the hand-eye measurements is lower than that of vector observations. This shows that the vector observations are effective in aiding the attitude determination with only hand-eye measurements. Also, one can obviously see that using symmetric $\bm{A}_i$ and $\bm{B}_i$, the accuracy is much lower than that in rigid ones. As in \cite{Modenini2018,Modenini2019} $\bm{A}_i, \bm{B}_i$ are symmetric, that is to say the method in Modenini's work can only achieve limited accuracy, regardless of its inevitable drawback in the angle observability. When combined with vector observations, the accuracy, robustness and observability can be dramatically improved. In the next sub-section, we are going to conduct real-world experiments using data from the Dawn spacecraft.

\begin{figure*}[hbp]
\centering
\includegraphics[width=1.0\textwidth]{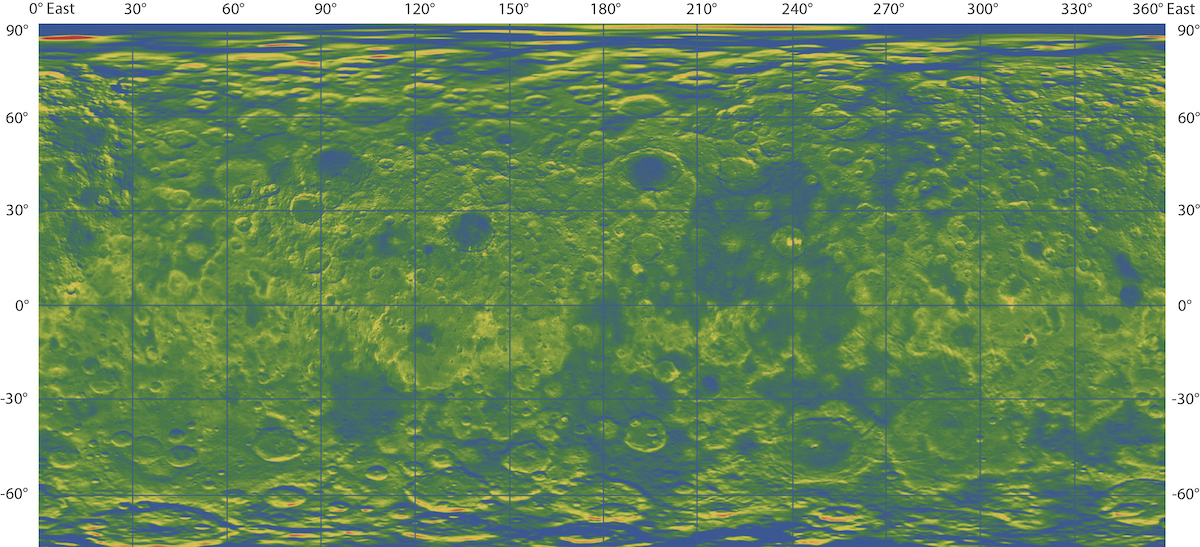}
\caption{The reconstructed terrain and altitude data using historical observations from the Dawn spacecraft. The darker color indicates lower altitude and vice versa.}
\label{fig:altitudes}
\end{figure*}

\subsection{Dawn Spacecraft Validation}
The Dawn spacecraft was launch by NASA on 27 Sept 2007 and has been aimed to discover the two dwarf planets i.e. Vesta and Ceres in the Kuiper belt. The Dawn mission ends on 1st Nov 2018 with the spacecraft consumed all the fuel for attitude control of its solar panel. The memorable Dawn mission had fullfiled the dreams of the scientists and enthusiasts for inspection of distant dwarf planets in the solar system. On the Dawn spacecraft, there were two framing cameras imaging the targets. Ceres has better geometric shape than Vesta since it is highly ellipsoidal and owns the semi-major axis of 482 km and semi-minor axis of 446 km. Therefore, in the study of Modenini \cite{Modenini2018,Modenini2019}, it has been successfully proven that such ellipsoid imaged in the camera frame can be used for accurate spacecraft attitude determination.\\
\indent As described before, the visual measurements can also provide attitude information from another aspect. Therefore, we verify the two sources of them i.e. the PnP and epipolar geometry in this sub-section. There are many representative point feature extraction methods proposed previously and each algorithm has its pros and cons. For the motion estimation using Ceres images, we choose the speed up robust features (SURF, \cite{bay2008speeded}), binary robust invariant scalable keypoints (BRISK, \cite{Leutenegger2011BRISK}), oriented FAST and rotated BRIEF (ORB, \cite{rublee2011orb}) for comparisons. The conic detection and the elliptic fittings are conducted according to \cite{Fitzgibbon1999,Quan1996,Zhang1997}.\\
\indent Using the epipolar geometry, the camera poses can be restored by integrating all sequential relative rotation matrices between successive images. Fig. \ref{fig:brisk}, \ref{fig:surf}, \ref{fig:orb} present the feature matching using BRISK, SURF and ORB respectively, for the images \texttt{FC21B0088753\_17042144605F2C} and \texttt{FC21B0088785\_17042150857F2C}. Here, the random sample consensus (RANSAC, \cite{chum2008optimal,Rawashdeh2014}) is invoked for figuring out the correspondences between 2D-2D image feature points. All the algorithms in this sub-section are implemented using the MATLAB computer vision toolbox. The feature numbers, the RANSAC successfull ratio and the execution time of various algorithms are summarized in Table \ref{tab:features}.

\begin{table}[h]
\centering
\caption{Stats of Various Feature Extraction Algorithms}
\label{tab:features}
\resizebox{0.5\textwidth}{!}{
\begin{tabular}{cccc}
\toprule
{Algorithms}&{Feature Numbers}&{RANSAC Successful Ratio}&{Time (s)}\\
\midrule
{BRISK}&{$84$}&{$97.62\%$}&{0.885}\\
{SURF}&{$237$}&{$84.38\%$}&{1.277}\\
{ORB}&{$24$}&{$100\%$}&{1.092}\\
\bottomrule
\end{tabular}}
\end{table}
\indent From the statistics, we can see that ORB extracts too few features but the RANSAC matching is the most successful. While the SURF consumes the most computation time and least RANSAC successful ratio (see the red markers in Fig. \ref{fig:surf}), the BRISK maintains a balance among the three ones. Therefore the BRISK feature descriptor is ultilized for the attitude determination validation. With the PnP method, a nonlinear semidefinite programming is employed to solve the 2D-3D registration problem between 2D image and the 3D terrain map \cite{Khoo2016}. Here the 3D terrain map is generated using the images of the Dawn spacecraft during the \texttt{Survey} mission of the Ceres, which can be found out at \texttt{https://sbib.psi.edu}. Fig. \ref{fig:altitudes} shows the exact details of such 3D terrain map.\\
\indent The reference attitude information from the Dawn spacecraft attitude estimator can be acquired from \texttt{LBL} files in the data folder. The real-world images from the Dawn spacecraft during the \texttt{GRaND} mission are grabbed for the verification of the proposed algorithm since in that image sequence the illuminance and the shape of the imaged Ceres are more appropriate than that in other datasets. The relative weighting of the vector observations and hand-eye measurements is tuned to $\varrho = 0.5$ indicating that the hand-eye measurements own twice accuracy than a single pair of feature correspondence. The PnP and epipolar geometry do not give 3D vector observation pairs directly. Instead, they are nonlinear and only the rotation matrices can be estimated. Suppose we have estimated the attitude matrices from PnP and epipolar geometry i.e. $\bm{R}_{\rm{PnP}}, \bm{R}_{\rm{epipolar}}$, with the similar approach in (\ref{glnr}), we can reconstruct the vector observations by
\begin{equation}
\left\{ {\begin{array}{*{20}{c}}
  {{{\bm{d}}_{{\rm{PnP}},1}} = {\bm{R}}_{\rm{PnP}}{{\bm{e}}_1}} \\ 
  {{{\bm{d}}_{{\rm{PnP}},2}} = {\bm{R}}_{\rm{PnP}}{{\bm{e}}_2}} \\ 
  {{{\bm{d}}_{{\rm{PnP}},3}} = {\bm{R}}_{\rm{PnP}}{{\bm{e}}_3}} \\ 
\end{array}} \right., \left\{ {\begin{array}{*{20}{c}}
  {{{\bm{d}}_{{\rm{epipolar}},1}} = {\bm{R}}_{\rm{epipolar}}{{\bm{e}}_1}} \\ 
  {{{\bm{d}}_{{\rm{epipolar}},2}} = {\bm{R}}_{\rm{epipolar}}{{\bm{e}}_2}} \\ 
  {{{\bm{d}}_{{\rm{epipolar}},3}} = {\bm{R}}_{\rm{epipolar}}{{\bm{e}}_3}} \\ 
\end{array}} \right.
\end{equation}
\noindent The reason for such reconstruction is two-fold: 1) If we directly solve hand-eye, PnP and epipolar geometry problems together, there will be too many local minima involved in the global searching for the highly non-convex combined new problem; 2) Due the local minima, estimating a controllable and accurate covariance description may be very tough and trivial. Such reconstruction of vectors will not loose accuracy of the obtained rotation since the rotation is exactly the optimal one on $SO(n) $corresponding to the above vector-matching problem \cite{Shuster1981,Bar-Itzhack2000}. For the epipolar geometry, the rotation matrices are propagated sequentially based on the estimates in the latest time instant. Then the attitude determination results are processed with the proposed method and previous representatives. Table \ref{tab:accuracy} consists of the root mean squared (RMS) statistics of the attitude errors in Euler angles.
\begin{table}[h]
\centering
\caption{RMS Accuracy Performances of Various Algorithms}
\label{tab:accuracy}
\resizebox{0.5\textwidth}{!}{
\begin{tabular}{cccc}
\toprule
{Algorithms}&{Roll (deg)}&{Pitch (deg)}&{Yaw (deg)}\\
\midrule
{Proposed-PnP}&{$5.33 \times 10^{-03}$}&{$3.21 \times 10^{-03}$}&{$5.81 \times 10^{-02}$}\\
{Proposed-Epipolar}&{$8.54 \times 10^{-03}$}&{$7.46 \times 10^{-03}$}&{$3.11 \times 10^{-02}$}\\
{PnP-Only}&{$1.22 \times 10^{-03}$}&{$3.01 \times 10^{-03}$}&{$3.78 \times 10^{-02}$}\\
{Modenini \cite{Modenini2018}}&{$8.21 \times 10^{-03}$}&{$6.04 \times 10^{-03}$}&{$1.99 \times 10^{+00}$}\\
\bottomrule
\end{tabular}}
\end{table}
\noindent It is clearly presented that the original algorithm by Modenini can give very accurate estimates of roll and pitch but does not have good performance in estimating the yaw angles. The reason has been shown in \cite{Modenini2018,Modenini2019} that for some perspective views the imaged ellipsoids are very close to spheroids which limits the observability on the yaw direction. The reason is not algorithmical but very physical since in many cases the flattenings of the imaged ellipsoid are small. When introducing the PnP and the epipolar geometry, the accuracy for yaw angles is significantly improved. In fact, the proposed method solves the attitude determination problem between various weighted measurements. The reason that PnP achieves better yaw estimation performance is that the imaging direction well corresponds to the yaw (see Fig. \ref{fig:brisk}). So the point feature correspondences significantly improve the the yaw accuracy. It also deserves a notification that the epipolar geometry does not achieve better roll and pitch precisions since the attitude from such source is sequentially integrated and thus it may suffer from low sampling frequency. However, epipolar geometry does not need to have preliminaries on the 3D model of the imaged ellipsoid. Therefore, from this aspect, epipolar geometry is more flexible than PnP. If the 3D terrain of the imaged ellipsoid is already known e.g. the Moon, then PnP should definitely replace the epipolar geometry for better accuracy and robustness.

\begin{figure}[h]
\centering
\includegraphics[width=0.45\textwidth]{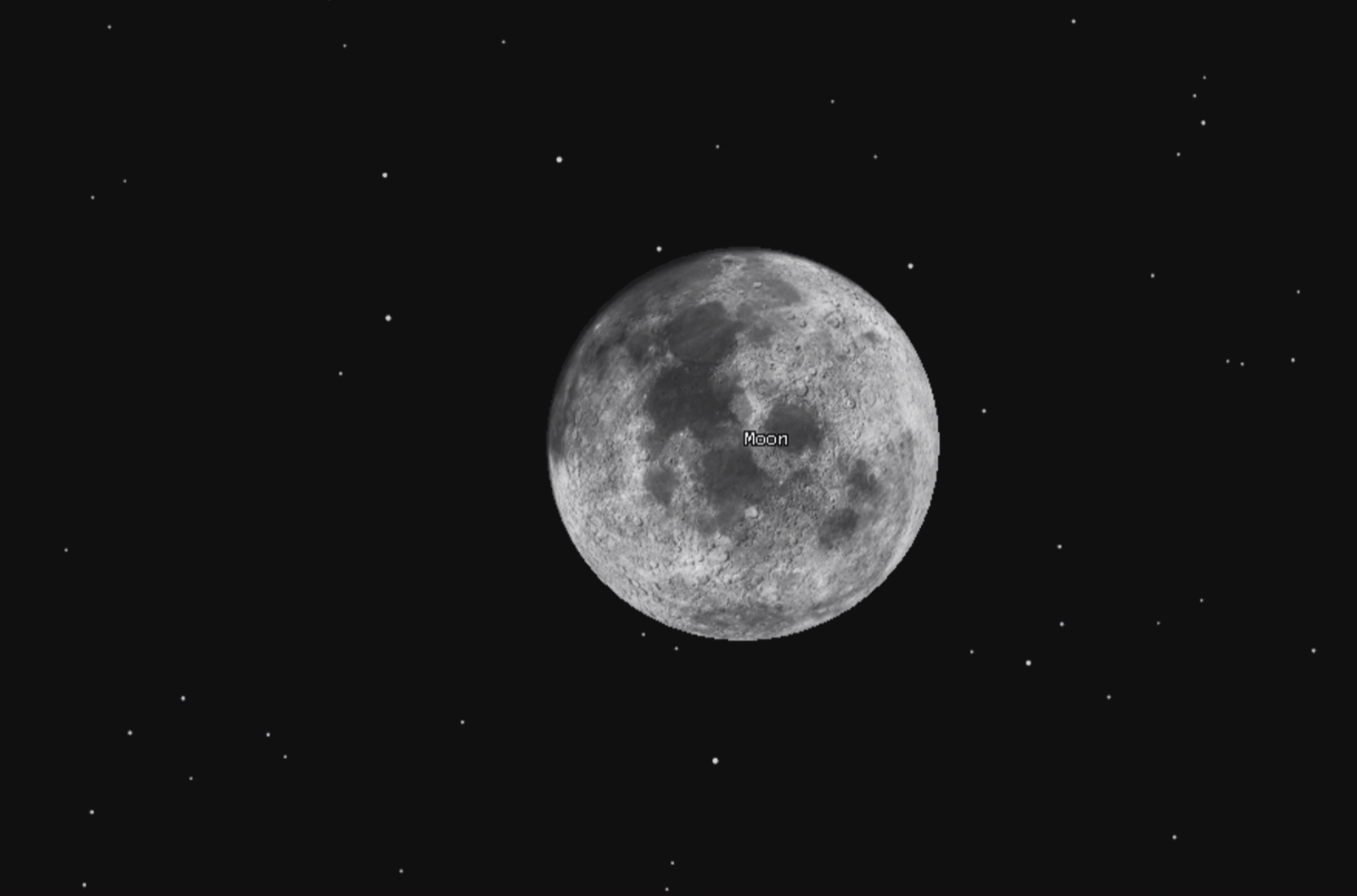}
\ \\
\ \\
%
\includegraphics[width=0.45\textwidth]{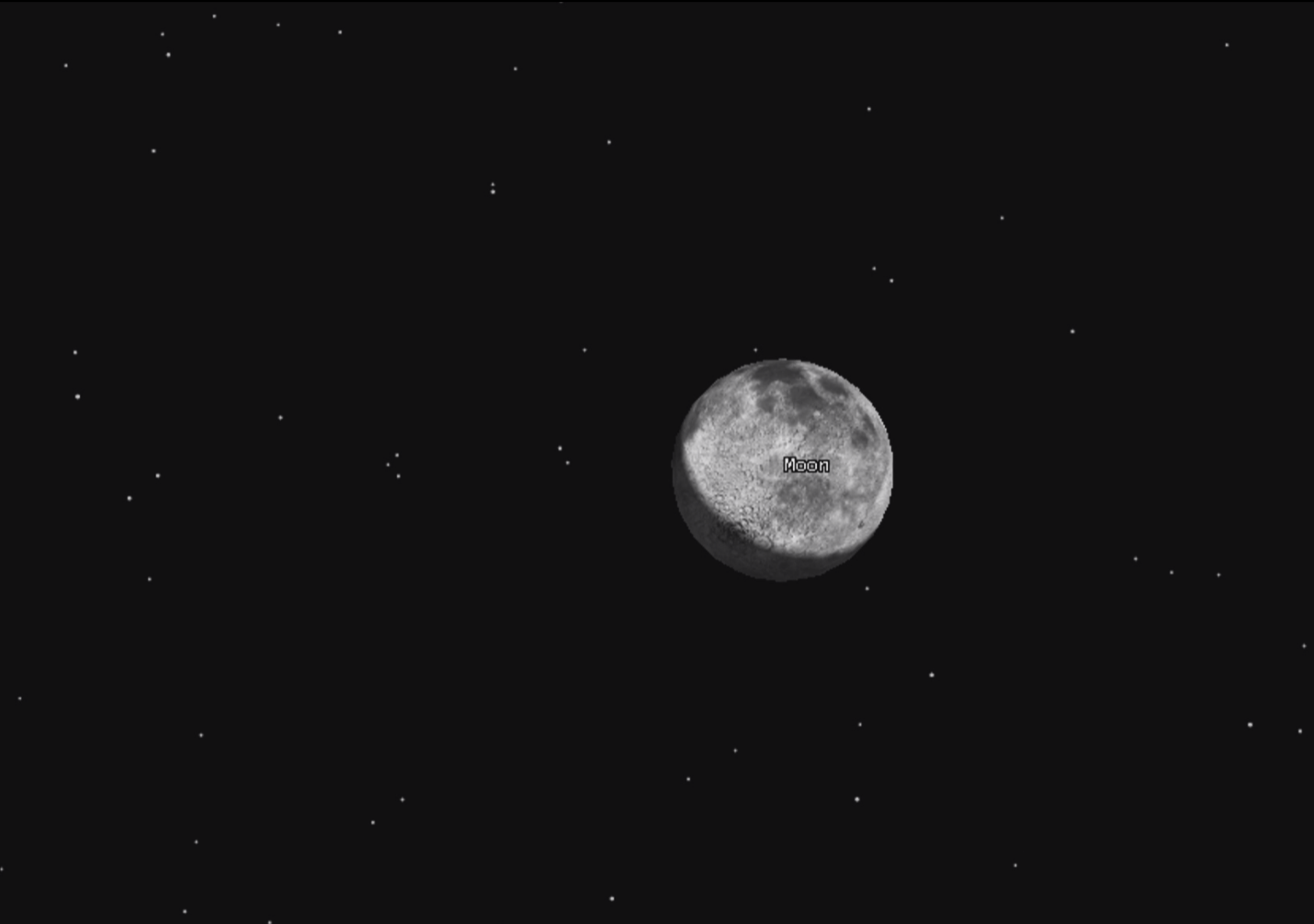}
\caption{The captured Lunar images during the simulated flight.}
\label{fig:moon2}
\end{figure}

\subsection{Integrating with Angular Rate}
In this sub-section, we simulate a case in the presence of angular rate measurements. Angular rate measurements mainly come from the onboard gyroscope or the estimation from star trackers, which are accurate and reliable devices. These measurements will provide two advances for the attitude determination scheme proposed in this paper:
\begin{enumerate}
\item They will largely improve the attitude update frequency.
\item They will significantly improve the attitude estimation accuracy.
\end{enumerate}
A spacecraft carrying high definition camera and gyroscope is modeled, whose inertia matrix is ${\mathcal{I}} = 4500\ {\rm{kg \cdot m^2}} \bm{I}$. The gyroscope contains unknown constant bias term to be estimated in further filter loops. We use the Moon as the central body and the orbit propagation is conducted via the high-precision orbit propagator (HPOP) from Jet Propulsion Laboratory (JPL). The captured Moon in different perspectives will be shown in Fig. \ref{fig:moon2}. The Lunar flattening is quite small, making it almost a perfect sphere with semi-major axis of 3,476.2 km and semi-minor axis of 3,472.0 km. Therefore, the observability of yaw is completely lost according to Modenini's theory \cite{Modenini2018}. Still, the vector observations are reconstructed from the rotation estimation using the epipolar geometry as described in the last sub-section. We use the Kalman filters for the state estimation. The preset attitude profile has been generated using the following unit quaternion model
\begin{equation}
\begin{gathered}
\bm{q}_i = \sin (i T \bm{\phi} + \bm{\psi}), \ \ \ \ \ i = 1, 2, \cdots, 20000 \hfill \\
\bm{q}_i = \bm{q}_i / \left \| \bm{q}_i \right \| \hfill \\
\end{gathered}
\end{equation}
with $T =  1 / (1000 \ \rm{Hz})$ being the sampling period and
\begin{equation}
\begin{gathered}
\bm{\phi} = (-0.12352, -0.31294, 0.62993, -0.27127)^\top \\
\bm{\psi} = (-0.74532, -0.24811, 0.66610, -0.54501)^\top \\
\end{gathered}
\end{equation}
The state vector is $\tilde{\bm{x}} = \left( \bm{x}^\top, \bm{\omega}^\top \right)^\top$ where $\bm{\omega}$ is the angular rate vector that owns the model of
\begin{equation}
\bm{\omega} = \hat{\bm{\omega}} + \bm{\omega}_{\rm{bias}} + \bm{\omega}_{\rm{noise}} 
\end{equation}
in which $\bm{\omega}_{\rm{bias}} = \left( \omega_{x, {\rm{bias}}}, \omega_{y, {\rm{bias}}}, \omega_{z, {\rm{bias}}}\right)^\top$ is the bias term and $\bm{\omega}_{\rm{noise}}$ denotes the noise term such that $\bm{\omega}_{\rm{noise}}  \sim {\mathcal{N}} (\bm{0}, \sigma_{\bm{\omega}}^2 \bm{I})$ with isotropic standard deviation of $\sigma_{\bm{\omega}}$. The process model of the Kalman filter is simply 
\begin{equation}
\left \{
\begin{gathered}
\mat(\dot{\bm{x}}) = - \bm{\omega}_{\times} \mat(\bm{x}) \hfill \\
\dot{\bm{\omega}}_{\rm{bias}} = \bm{0} \hfill \\
\end{gathered}
\right.
\end{equation}
where $\bm{\omega}_{\times}$ is the skew-symmetric matrix of $\bm{\omega}$. The measurement model is
\begin{equation}
\bm{y} = \bm{K} \tilde{\bm{x}}
\end{equation} 
where $\bm{y} = \left( \bm{x}_{\rm{meas}}^\top, \bm{0} \right)^\top \in \mathbb{R}^{12}$ is the measurement vector and the measurement matrix $\bm{K} \in \mathbb{R}^{12 \times 12}$ is
\begin{equation}
\bm{K} = \left( 
\begin{gathered}
\bm{I}\ \ \ \bm{0} \\
\bm{0}\ \ \ \bm{0} \\
\end{gathered}
\right)
\end{equation}
Here $\bm{x}_{\rm{meas}}$ denotes the computed measurement rotation matrix using proposed method from (\ref{x_sol}). Since here the state $\bm{x}$ is constrained on $SO(n)$, a recent filter with such constraint has been invoked \cite{Ruiter2017}. The extended Kalman filter (EKF) and unscented Kalman filter (UKF) mechanisms are employed to \cite{Ruiter2017} for state estimation. The related parameters are
\begin{enumerate}
\item Initial state: $\bm{\alpha}_0 = \left[ \vecc(\bm{I}), \bm{0} \right]^\top$.
\item Initial state covariance: $\bm{\Sigma}_{\tilde{\bm{x}}_0} = 1 \times 10^{-1} \bm{I}$.
\item Gyroscope standard deviation: $\sigma_{\bm{\omega}} = 1 \times 10^{-2} \ {\rm{rad/s}}$. 
\item Bias standard deviation: $\sigma_{\bm{\omega}_{\rm{bias}}} = 1 \times 10^{-3} \ {\rm{rad/s}}$. 
\item Measurement covariance: $\bm{\Sigma}_{\bm{y}} = \left( 
\begin{gathered}
\bm{\Sigma}_{\bm{x}\bm{x}}\ \ \ \bm{0} \\
\bm{0}\ \ \ \ \ \bm{0} \\
\end{gathered}
\right)$.
\end{enumerate}

\begin{figure}[h]
\centering
\includegraphics[width=0.5\textwidth]{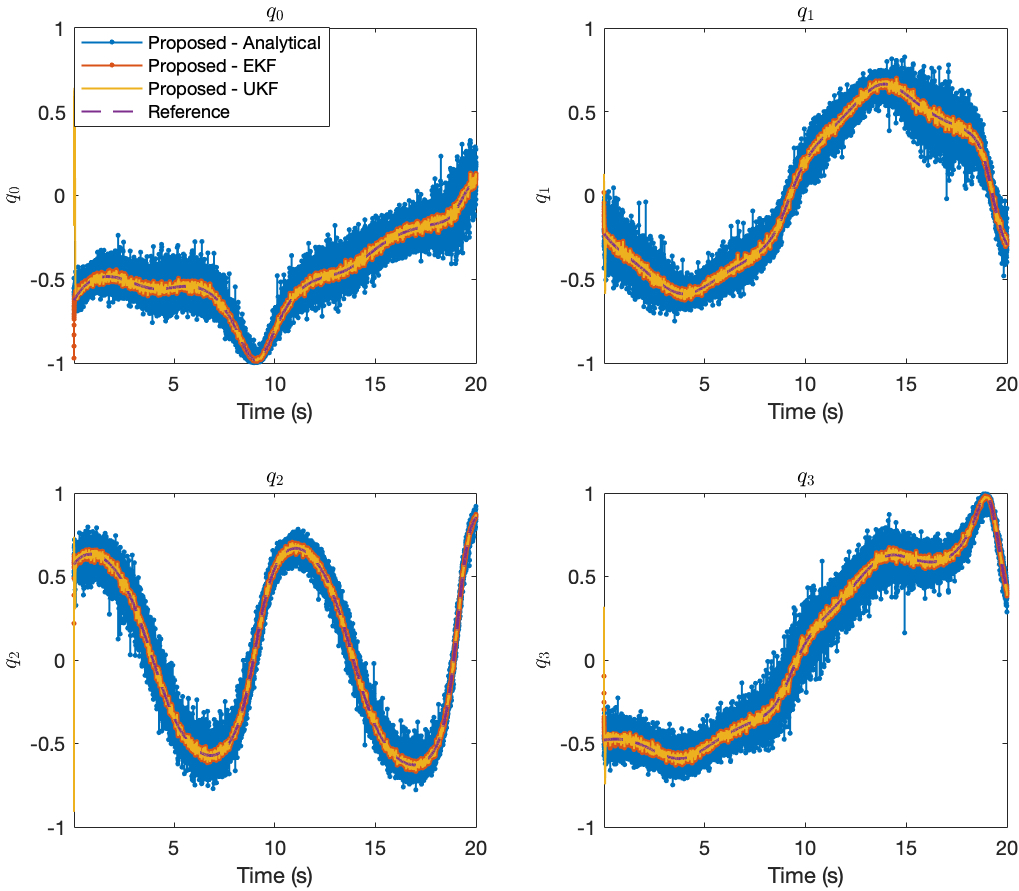}
\caption{Attitude estimation results using the angular rate, epipolar geometry and hand-eye measurements.}
\label{fig:quat_gyro}
\ \\
\ \\
%
\centering
\includegraphics[width=0.5\textwidth]{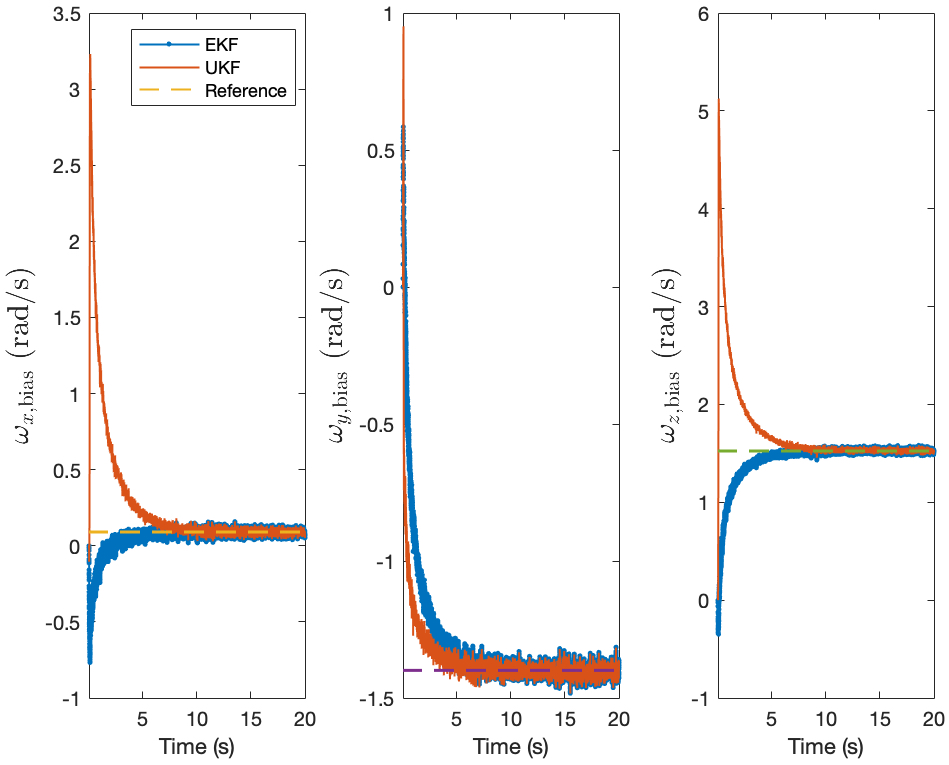}
\caption{Estimated gyroscope biases.}
\label{fig:gyro_bias}
\end{figure}

\noindent The state estimation and proposed algorithms are implemented on an embedded STM32F407VG chip with clock speed of 168MHz. The feature extraction has been conducted using the BRISK descriptor on a field programmable gate array (FPGA) guaranteeing a stable processing image processing speed of 120fps. The predicting loop is performed at 1000Hz and the measurement correction is deployed at 120Hz in accordance with the feature extraction. The codes are programmed using the C++11 standard. The attitude estimation results in the form of quaternion from various algorithms are shown in Fig. \ref{fig:quat_gyro}. The proposed analytical results can follow the reference but the noise scale is quite large. When the Kalman filter is introduced for fusion with gyroscope, the estimation errors significantly decrease, leading to accurate estimation of the gyroscope biases presented in Fig. \ref{fig:gyro_bias}. Both EKF and UKF can converge to the correct reference values and EKF converges a little bit slower than UKF. This verifies that when combined with inertial measurements, the proposed method along with its covariance information are accurate and reliable. The load of the central processing unit (CPU) has been summarized via the internal scheduling unit of the embedded operating system. Table \ref{tab:stats}. From the presented stats, we are able to see that the proposed method together with its Kalman filter version with inertial measurements are computationally efficient for execution on the embedded platform. 

\begin{table}[h]
\centering
\caption{Computational CPU Load (Averaged)}
\label{tab:stats}
\resizebox{0.3\textwidth}{!}{
\begin{tabular}{cc}
\toprule
{Algorithms}&{CPU Load)}\\
\midrule
{Proposed - No Covariance}&{$3.986\%$}\\
{Proposed - With Covariance}&{$6.710\%$}\\
{Proposed - EKF}&{$27.043\%$}\\
{Proposed - UKF}&{$34.925\%$}\\
\bottomrule
\end{tabular}}
\end{table}

\section{Conclusion}
The original attitude determination using hand-eye measurements derived from images of ellipsoid has limited accuracy, robustness and observability. Vector observations can well aid the spacecraft attitude determination with hand-eye measurements and thus enhance the overall performances. It is noted that the vector observations can come from various sources so the proposed scheme is quite universal. These aspects contribute to a widened approach for visual attitude determination for spacecraft in deep spaces. The experimental studies have verified the effectiveness of the proposed scheme for in-flight attitude determination. The shown results indicates that the proposed combined approach is accurate for Dawn spacecraft. A further synthetic study on the fusion with inertial measurements has also been conducted showing the flexibility and superiority of multiple-source integration. \\
\indent The presented approach also has its drawback i.e. the estimated $\bm{x}$ is not strictly under the constraint of $SO(n)$ and requires a further orthonormalization operation. The next task for us is to find a fully $SO(n)$-constrained solution on Lie groups. Besides, the weighting strategy in this paper is currently empirical. And how to optimally determine a best set of weights in real-time will also be a challenging task.\\
\indent For low-cost purposes, the ellipsoidal markers can be installed onto spacecrafts for relative attitude determination using visual correspondences and hand-eye measurements. Also, in the total design of the Dawn spacecraft, there are two framing cameras mounted targeting to the same direction. This will motivate us to invoke the stereo imaging principles to give more accurate attitude determination results in the future.

\begin{figure*}[hb]
\hrulefill
\begin{equation}\label{deriv_1}
\begin{small}
\begin{gathered}
\bm{S}_2 = \hfill \\
\ \ \ \ \left<
\begin{gathered}
\left[\delta \bm{H}  + \left(\delta\bm{Q} \otimes \bm{I}\right)\left(\bm{Q}^{\top} \otimes \bm{I}\right) + \left(\bm{Q} \otimes \bm{I}\right)\left(\delta\bm{Q}^{\top} \otimes \bm{I}\right) \right] \bm{x}\bm{x}^{\top}\left[\delta \bm{H}  + \left(\delta\bm{Q} \otimes \bm{I}\right)\left(\bm{Q}^{\top} \otimes \bm{I}\right) + \left(\bm{Q} \otimes \bm{I}\right)\left(\delta\bm{Q}^{\top} \otimes \bm{I}\right) \right]^{\top}
\end{gathered}
\right> \hfill \\
\approx \left<
\delta \bm{H} \bm{x}\bm{x}^{\top}\delta \bm{H} + \left[\left(\delta\bm{Q} \otimes \bm{I}\right)\left(\bm{Q}^{\top} \otimes \bm{I}\right) + \left(\bm{Q} \otimes \bm{I}\right)\left(\delta\bm{Q}^{\top} \otimes \bm{I}\right)\right]\bm{x}\bm{x}^{\top}\left[\left(\delta\bm{Q} \otimes \bm{I}\right)\left(\bm{Q}^{\top} \otimes \bm{I}\right) + \left(\bm{Q} \otimes \bm{I}\right)\left(\delta\bm{Q}^{\top} \otimes \bm{I}\right)\right]^{\top}
\right> \hfill \\
= \underbrace{\left( \sum \limits_{i = 1}^{M}  v_i 
 {\mathcal{F}} \left[\left(\bm{A}_i \otimes \bm{I} - \bm{I} \otimes \bm{B}_i^{\top} \right)\bm{x}\right] 
\right)}_{n^2 \times 2n^2} \bm{\Sigma}_{\bm{m}_i} \underbrace{\left( \sum \limits_{i = 1}^{M}  v_i 
 {\mathcal{F}} \left[\left(\bm{A}_i \otimes \bm{I} - \bm{I} \otimes \bm{B}_i^{\top} \right)\bm{x}\right] 
\right)^{\top}}_{2n^2 \times n^2}  \hfill \\
\ \ \ \ \ + \left[ \sum \limits_{i = 1}^{M}  v_i 
 \underbrace{\left(\bm{A}_i^{\top} \otimes \bm{I} - \bm{I} \otimes \bm{B}_i \right)}_{n^2 \times n^2} \underbrace{{\mathcal{F}}(\bm{x})}_{n^2 \times 2n^2}
\right] \bm{\Sigma}_{\bm{n}_i} \left[ \sum \limits_{i = 1}^{M}  v_i 
  \left(\bm{A}_i^{\top} \otimes \bm{I} - \bm{I} \otimes \bm{B}_i \right){\mathcal{F}}(\bm{x})
\right]^{\top} \hfill \\
\ \ \ \ \ + {\mathcal{Z}}\left( \bm{RQ} \right)\bm{\Sigma}_{\vecc\left( \bm{Q}^{\top} \right)} {\mathcal{Z}}^{\top}\left( \bm{RQ} \right) + \left( \bm{Q} \otimes \bm{R} \right)\bm{\Sigma}_{\vecc\left( \bm{Q}\right)} \left( \bm{Q}^{\top} \otimes \bm{R}^{\top} \right) \hfill \\
\ \ \ \ \ + {\mathcal{Z}}\left( \bm{RQ} \right)\bm{\Sigma}_{\vecc\left( \bm{Q}^{\top} \right), \vecc(\bm{Q})} \left( \bm{Q}^{\top} \otimes \bm{R}^{\top} \right) + \left( \bm{Q} \otimes \bm{R} \right) \bm{\Sigma}_{\vecc\left( \bm{Q} \right), \vecc \left(\bm{Q}^{\top} \right)} {\mathcal{Z}}^{\top}\left( \bm{RQ} \right) \hfill \\
\end{gathered}
\end{small}
\end{equation}
\hrulefill
\begin{equation}\label{deriv_2}
\begin{gathered}
\bm{S}_3 = \hfill \\
\ \ \ - \left<
\left[ \left(\delta\bm{Q} \otimes \bm{I}\right)\vecc(\bm{P}) + \left(\bm{Q} \otimes \bm{I}\right)\vecc(\delta \bm{P})\right]\bm{x}^{\top}\left[\delta \bm{H}  + \left(\delta\bm{Q} \otimes \bm{I}\right)\left(\bm{Q}^{\top} \otimes \bm{I}\right) + \left(\bm{Q} \otimes \bm{I}\right)\left(\delta\bm{Q}^{\top} \otimes \bm{I}\right) \right]^{\top}
\right>\hfill \\
\approx - \left<
\left[ {\mathcal{Z}}(\bm{P}) \vecc(\delta \bm{Q}^{\top})  + \left(\bm{Q} \otimes \bm{I}\right)\vecc(\delta \bm{P})\right]\left[{\mathcal{Z}}\left( \bm{RQ} \right)\vecc\left(\delta \bm{Q}^{\top}\right) + \left(\bm{Q} \otimes \bm{R}\right) \vecc(\delta \bm{Q})\right]^{\top}
\right> \hfill \\
= - {\mathcal{Z}}\left( \bm{P} \right)\bm{\Sigma}_{\vecc\left( \bm{Q}^{\top} \right)} {\mathcal{Z}}^{\top}\left( \bm{RQ} \right) - \left( \bm{Q} \otimes \bm{I} \right) \bm{\Sigma}_{\vecc(\bm{P}), \vecc(\bm{Q})}\left( \bm{Q}^{\top} \otimes \bm{R}^{\top} \right) \hfill \\
\ \ \ - {\mathcal{Z}}\left( \bm{P} \right)\bm{\Sigma}_{\vecc \left(\bm{Q}^{\top} \right), \vecc(\bm{Q})}\left( \bm{Q}^{\top} \otimes \bm{R}^{\top} \right) - \left( \bm{Q} \otimes \bm{I} \right) \bm{\Sigma}_{\vecc(\bm{P}), \vecc \left(\bm{Q}^{\top} \right)}{\mathcal{Z}}^{\top}\left( \bm{RQ} \right) \hfill \\
\end{gathered}
\end{equation}
\end{figure*}


%

\appendices
\section{Matrix Derivations}
The items inside the internal expectation of $\bm{\Sigma}_{\bm{xx}}$ can be derived as follows
\begin{equation}
\begin{gathered}
\left[ \left(\delta\bm{Q} \otimes \bm{I}\right)\vecc(\bm{P}) + \left(\bm{Q} \otimes \bm{I}\right)\vecc(\delta \bm{P})\right] \hfill \\
\ \ \ \ \ \left[ \left(\delta\bm{Q} \otimes \bm{I}\right)\vecc(\bm{P}) + \left(\bm{Q} \otimes \bm{I}\right)\vecc(\delta \bm{P})\right] ^{\top} \hfill \\
= \left[ \left(\delta\bm{Q} \otimes \bm{I}\right)\vecc(\bm{P}) + \left(\bm{Q} \otimes \bm{I}\right)\vecc(\delta \bm{P})\right]  \hfill \\
\ \ \ \ \ \left[ \vecc(\bm{P})^{\top}\left(\delta\bm{Q}^{\top} \otimes \bm{I}\right) + \vecc(\delta \bm{P})^{\top}\left(\bm{Q}^{\top} \otimes \bm{I}\right)\right] \hfill \\
= \left(\delta\bm{Q} \otimes \bm{I}\right)\vecc(\bm{P})\vecc(\bm{P})^{\top}\left(\delta\bm{Q}^{\top} \otimes \bm{I}\right) + \hfill \\ 
\ \ \ \ \ \left(\delta\bm{Q} \otimes \bm{I}\right)\vecc(\bm{P})\vecc(\delta \bm{P})^{\top}\left(\bm{Q}^{\top} \otimes \bm{I}\right) + \hfill \\
\ \ \ \ \ \left(\bm{Q} \otimes \bm{I}\right)\vecc(\delta \bm{P})\vecc(\bm{P})^{\top}\left(\delta\bm{Q}^{\top} \otimes \bm{I}\right) + \hfill \\ 
\ \ \ \ \ \left(\bm{Q} \otimes \bm{I}\right)\vecc(\delta \bm{P})\vecc(\delta \bm{P})^{\top}\left(\bm{Q}^{\top} \otimes \bm{I}\right) \hfill \\
%
\Rightarrow \bm{S}_1 = \left<
\begin{gathered}
\left[ \left(\delta\bm{Q} \otimes \bm{I}\right)\vecc(\bm{P}) + \left(\bm{Q} \otimes \bm{I}\right)\vecc(\delta \bm{P})\right] \\
\left[ \left(\delta\bm{Q} \otimes \bm{I}\right)\vecc(\bm{P}) + \left(\bm{Q} \otimes \bm{I}\right)\vecc(\delta \bm{P})\right] ^{\top} 
\end{gathered}
\right> \hfill \\
= \left<
\begin{gathered}
{\mathcal{Z}}(\bm{P}) \vecc(\delta \bm{Q}^{\top}) \vecc(\delta \bm{Q}^{\top})^{\top}{\mathcal{Z}}^{\top}(\bm{P}) + \hfill \\
{\mathcal{Z}}(\bm{P}) \vecc(\delta \bm{Q}^{\top})\vecc(\delta \bm{P})^{\top}\left(\bm{Q}^{\top} \otimes \bm{I}\right) + \hfill \\
\left(\bm{Q} \otimes \bm{I}\right)\vecc(\delta \bm{P})\vecc(\delta \bm{Q}^{\top})^{\top}{\mathcal{Z}}^{\top}(\bm{P}) + \hfill \\
\left(\bm{Q} \otimes \bm{I}\right)\vecc(\delta \bm{P})\vecc(\delta \bm{P})^{\top}\left(\bm{Q}^{\top} \otimes \bm{I}\right) \hfill \\
\end{gathered}
\right>\hfill \\
= {\mathcal{Z}}(\bm{P}){\bm{\Sigma}}_{\vecc(\bm{Q}^\top)}{\mathcal{Z}}^{\top}(\bm{P}) \hfill \\
\ \ \ \ \ + {\mathcal{Z}}(\bm{P})\bm{\Sigma}_{\vecc\left( \bm{Q}^{\top}\right), \vecc(\bm{P})} \left(\bm{Q}^{\top} \otimes \bm{I}\right)\hfill \\
\ \ \ \ \ + \left(\bm{Q} \otimes \bm{I}\right) \bm{\Sigma}_{\vecc(\bm{P}), \vecc\left( \bm{Q}^{\top}\right)} {\mathcal{Z}}^{\top}(\bm{P}) \hfill \\
\ \ \ \ \ + \left(\bm{Q} \otimes \bm{I}\right){\bm{\Sigma}}_{\vecc(\bm{P})} \left(\bm{Q}^{\top} \otimes \bm{I}\right)\hfill \\
\end{gathered}
\end{equation}
Denoting
\begin{equation}
\begin{gathered}
\bm{\Sigma}_{\bm{m}_i} = \underbrace{\begin{array}{l}
\left[ {\begin{array}{*{20}{c}}
{\bm{\Sigma}_{\vecc(\bm{A}_i)}}&{\bm{\Sigma}_{\vecc(\bm{A}_i), \vecc(\bm{B}_i)}}\\
{\bm{\Sigma}_{\vecc(\bm{B}_i), \vecc(\bm{A}_i)}}&{\bm{\Sigma}_{\vecc(\bm{B}_i)}}
\end{array}} \right]
\end{array}}_{2n^2 \times 2n^2} \hfill \\
\bm{\Sigma}_{\bm{n}_i} = \underbrace{\begin{array}{l}
\left[ {\begin{array}{*{20}{c}}
{\bm{\Sigma}_{\vecc(\bm{A}_i^{\top})}}&{\bm{\Sigma}_{\vecc(\bm{A}_i^{\top}), \vecc(\bm{B}_i^{\top})}}\\
{\bm{\Sigma}_{\vecc(\bm{B}_i^{\top}), \vecc(\bm{A}_i^{\top})}}&{\bm{\Sigma}_{\vecc(\bm{B}_i^{\top})}}
\end{array}} \right]
\end{array}}_{2n^2 \times 2n^2} \hfill \\
\end{gathered}
\end{equation}
and invoking
\begin{equation}
\begin{gathered}
 \left(\bm{Q} \otimes \bm{I}\right)\left(\delta\bm{Q}^{\top} \otimes \bm{I}\right)\bm{x} =  \left(\bm{Q} \otimes \bm{I}\right) \vecc \left( \bm{R} \delta \bm{Q}\right)\hfill \\
 = \left(\bm{Q} \otimes \bm{I}\right) \left(\bm{I} \otimes \bm{R}\right) \vecc(\delta \bm{Q}) \hfill \\
 =  \left(\bm{Q} \otimes \bm{R}\right) \vecc(\delta \bm{Q}) \hfill \\
 \end{gathered}
\end{equation}
we obtain $\bm{S}_2, \bm{S}_3$ in (\ref{deriv_1}) and (\ref{deriv_2}). Note that in these derivations, the cross correlation between vector observations and hand-eye measurements is ignored as they come from completely different sources so a cross-correlation evaluation may be trivial. Since we have

\begin{equation}
\begin{gathered}
\bm{P}^{\top} = \left(
\begin{gathered}
\sqrt{w_1}\bm{b}_1^{\top}  \\
\sqrt{w_2}\bm{b}_2^{\top}  \\
\vdots \\
\sqrt{w_N}\bm{b}_N^{\top}  \\
\end{gathered}
\right), \bm{Q}^{\top} = \left(
\begin{gathered}
\sqrt{w_1}\bm{r}_1^{\top}  \\
\sqrt{w_2}\bm{r}_2^{\top}  \\
\vdots \\
\sqrt{w_N}\bm{r}_N^{\top}  \\
\end{gathered}
\right)
\end{gathered}
\end{equation}
let
\begin{equation}
\bm{p}_i = \left(
\begin{gathered}
\sqrt{w_1}b_{1, i}  \\
\sqrt{w_2}b_{2, i}  \\
\vdots \\
\sqrt{w_N}b_{N, i}  \\
\end{gathered}
\right), \bm{q}_i = \left(
\begin{gathered}
\sqrt{w_1}r_{1, i}  \\
\sqrt{w_2}r_{2, i}  \\
\vdots \\
\sqrt{w_N}r_{N, i}  \\
\end{gathered}
\right), i = 1, 2, \cdots, n
\end{equation}
we have
\begin{equation}
\begin{gathered}
\bm{P}^{\top} = \left(\bm{p}_1, \bm{p}_2, \cdots, \bm{p}_n \right) \hfill \\
\bm{Q}^{\top} = \left(\bm{q}_1, \bm{q}_2, \cdots, \bm{q}_n \right) \hfill \\
\end{gathered}
\end{equation}

\begin{equation*}
\begin{gathered}
\bm{\Sigma}_{\vecc\left( \bm{P}^{\top}\right), \vecc(\bm{Q})} = \left<
\vecc\left( \delta \bm{P}^{\top}\right)  \vecc(\delta \bm{Q})^{\top}
\right> \hfill \\
= \left< 
\left(
\begin{gathered}
\delta \bm{p}_1 \\
\delta \bm{p}_2 \\
\vdots \\
\delta \bm{p}_n \\
\end{gathered}
\right) \left( \sqrt{w_1}\delta \bm{r}_1^{\top}, \sqrt{w_2}\delta \bm{r}_2^{\top}, \cdots, \sqrt{w_N}\delta \bm{r}_N^{\top}\right)
\right> \hfill \\
= \left< 
\left(
\begin{gathered}
\sqrt{w_1} \delta \bm{p}_1 \delta \bm{r}_1^{\top}, \sqrt{w_2} \delta \bm{p}_1 \delta \bm{r}_2^{\top}, \cdots, \sqrt{w_N} \delta \bm{p}_1 \delta \bm{r}_N^{\top} \hfill \\
\sqrt{w_1} \delta \bm{p}_2 \delta \bm{r}_1^{\top}, \sqrt{w_2} \delta \bm{p}_2 \delta \bm{r}_2^{\top}, \cdots, \sqrt{w_N} \delta \bm{p}_2 \delta \bm{r}_N^{\top} \hfill \\
\ \ \ \ \ \ \ \vdots \ \ \ \ \ \ \ \ \ \ \ \ \ \ \ \ \ \ \vdots \ \ \ \ \ \ \ddots \ \ \ \ \ \ \ \ \ \vdots \hfill \\
\sqrt{w_1} \delta \bm{p}_n \delta \bm{r}_1^{\top}, \sqrt{w_2} \delta \bm{p}_n \delta \bm{r}_2^{\top}, \cdots, \sqrt{w_N} \delta \bm{p}_n \delta \bm{r}_N^{\top} \hfill \\
\end{gathered}
\right)
\right>\hfill \\
\end{gathered}
\end{equation*}
\begin{equation}
\begin{gathered}
= \underbrace{\left(
\begin{gathered}
\sqrt{w_1} \bm{\Sigma}_{\bm{p}_1, \bm{r}_1}, \sqrt{w_2} \bm{\Sigma}_{\bm{p}_1, \bm{r}_2}, \cdots, \sqrt{w_N} \bm{\Sigma}_{\bm{p}_1, \bm{r}_N} \hfill \\
\sqrt{w_1} \bm{\Sigma}_{\bm{p}_2, \bm{r}_1}, \sqrt{w_2} \bm{\Sigma}_{\bm{p}_2, \bm{r}_2}, \cdots, \sqrt{w_N} \bm{\Sigma}_{\bm{p}_2, \bm{r}_N} \hfill \\
\ \ \ \ \ \ \ \vdots \ \ \ \ \ \ \ \ \ \ \ \ \ \ \ \vdots \ \ \ \ \ \ \ddots \ \ \ \ \ \ \ \ \ \vdots \hfill \\
\sqrt{w_1} \bm{\Sigma}_{\bm{p}_n, \bm{r}_1}, \sqrt{w_2} \bm{\Sigma}_{\bm{p}_n, \bm{r}_2}, \cdots, \sqrt{w_N} \bm{\Sigma}_{\bm{p}_n, \bm{r}_N} \hfill \\
\end{gathered}
\right)}_{nN \times nN}
\end{gathered}
\end{equation}
Then $\bm{\Sigma}_{\vecc\left( \bm{Q}^{\top}\right), \vecc(\bm{Q})}, \bm{\Sigma}_{\vecc\left( \bm{Q}^{\top}\right), \vecc(\bm{P})}, \bm{\Sigma}_{\vecc\left( \bm{P}\right), \vecc(\bm{Q}^{\top})},$ can be computed in the same manner. Note that if there is no auto-correlation between $\bm{b}_1, \bm{b}_2, \cdots, \bm{b}_N$, no auto-correlation between $\bm{r}_1, \bm{r}_2, \cdots, \bm{r}_N$ and no cross-correlation between $\bm{b}_i, i = 1, 2, \cdots, N$ and $\bm{r}_j, j = 1, 2, \cdots, N$, one arrives at
\begin{equation}
\begin{gathered}
\bm{\Sigma}_{\bm{p}_i, \bm{r}_j} = \bm{0} \\
\bm{\Sigma}_{\bm{q}_i, \bm{b}_j} = \bm{0} \\
\underbrace{\bm{\Sigma}_{\bm{p}_i, \bm{b}_j} = \left(
\begin{gathered}
\ddots \ \ \ \ \ \ \ \ \bm{0} \\
\sigma_{b_{j, i}}^2 \\
\bm{0}\ \ \ \ \ \ \ \ \ \ddots \\
\end{gathered}
\right)}_{N \times n}, \underbrace{\bm{\Sigma}_{\bm{q}_i, \bm{r}_j} = \left(
\begin{gathered}
\ddots \ \ \ \ \ \ \ \ \bm{0} \\
\sigma_{r_{j, i}}^2 \\
\bm{0}\ \ \ \ \ \ \ \ \ \ddots \\
\end{gathered}
\right)}_{N \times n}
 \hfill \\
\end{gathered}
\end{equation}
where $i = 1, 2, \cdots, n$, $j = 1, 2, \cdots, N$. The auto-covariance of $\vecc\left( \bm{P}^{\top}\right)$ is given by
\begin{equation*}
\begin{gathered}
\bm{\Sigma}_{\vecc\left( \bm{P}^{\top}\right)} = \left<
\vecc\left( \delta \bm{P}^{\top}\right)\vecc\left( \delta \bm{P}^{\top}\right)^{\top}
\right> \hfill \\
\end{gathered}
\end{equation*}
\begin{equation}
\begin{gathered}
= \left<
\left(
\begin{gathered}
\delta \bm{p}_1 \\
\delta \bm{p}_2 \\
\vdots \\
\delta \bm{p}_n \\
\end{gathered}
\right) \left(\delta \bm{p}_1^{\top}, \delta \bm{p}_2^{\top}, \cdots, \delta \bm{p}_n^{\top} \right)
\right> \hfill \\
= \left(
\begin{gathered}
\bm{\Sigma}_{\bm{p}_1, \bm{p}_1}, \bm{\Sigma}_{\bm{p}_1, \bm{p}_2}, \cdots, \bm{\Sigma}_{\bm{p}_1, \bm{p}_n} \hfill \\
\bm{\Sigma}_{\bm{p}_2, \bm{p}_1}, \bm{\Sigma}_{\bm{p}_2, \bm{p}_2}, \cdots, \bm{\Sigma}_{\bm{p}_2, \bm{p}_n} \hfill \\
\ \ \ \ \ \vdots \ \ \ \ \ \ \ \ \ \vdots \ \ \ \ddots \ \ \ \ \ \ \vdots \hfill \\
\bm{\Sigma}_{\bm{p}_n, \bm{p}_1}, \bm{\Sigma}_{\bm{p}_n, \bm{p}_2}, \cdots, \bm{\Sigma}_{\bm{p}_n, \bm{p}_n} \hfill \\
\end{gathered}
\right)
\end{gathered}
\end{equation}
and likewise $\bm{\Sigma}_{\vecc\left( \bm{Q}^{\top}\right)}$ can be obtained. If there is no auto-correlation between $\bm{b}_1, \bm{b}_2, \cdots, \bm{b}_N$ and no auto-correlation between $\bm{r}_1, \bm{r}_2, \cdots, \bm{r}_N$, we have
\begin{equation}
\begin{gathered}
\bm{\Sigma}_{\bm{p}_i, \bm{p}_j} =  \underbrace{\diag\left(
\sigma_{b_{1, i}, b_{1, j}}^2, \sigma_{b_{2, i}, b_{2, j}}^2, \cdots, \sigma_{b_{n, i}, b_{n, j}}^2
\right)}_{N \times N} \\
\bm{\Sigma}_{\bm{q}_i, \bm{q}_j} =  \underbrace{\diag\left(
\sigma_{r_{1, i}, r_{1, j}}^2, \sigma_{r_{2, i}, r_{2, j}}^2, \cdots, \sigma_{r_{n, i}, r_{n, j}}^2
\right)}_{N \times N} \\
\end{gathered}
\end{equation}

%

\section*{Acknowledgment}
The author would like to thank Dr. D. Modenini from University of Bologna, Italy for his constructive discussion on the hand-eye calibration and attitude determination problems. \\
\indent This research has been supported by National Natural Science Foundation of China under the grant of No. 41604025 and in part by the Open Project of Shanghai Key Laboratory of Navigation and Location-based Services, Shanghai Jiao Tong University.\\
\indent Some illustrative examples and codes of this paper are public and can be accessed at \textbf{\url{https://github.com/zarathustr/vec\_hand\_eye\_att}}.

\ifCLASSOPTIONcaptionsoff
  \newpage
\fi

\begin{IEEEbiography}[{\includegraphics[width=1in,height=1.25in,clip,keepaspectratio]{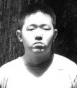}}]{Jin Wu} was born in May, 1994 in Zhenjiang, China. He received the B.S. Degree from University of Electronic Science and Technology of China, Chengdu, China. He has been a research assistant in Department of Electronic and Computer Engineering, Hong Kong University of Science and Technology since 2018. His research interests include robot navigation, multi-sensor fusion, automatic control and mechatronics. He is a co-author of over 40 technical papers in representative journals and conference proceedings of \textsc{IEEE, AIAA, IET} and etc. Mr. Jin Wu received the outstanding reviewer award for \textsc{Asian Journal of Control}. One of his papers published in \textsc{IEEE Transactions on Automation Science and Engineering} was selected as the ESI Highly Cited Paper by ISI Web of Science during 2017 to 2018. He is a member of IEEE.\\
\indent Mr. J. Wu started software programming using C++ since 2004. He has been in the UAV industry from 2012 and has launched two companies ever since. He has been a part-time Robotics Engineer in Tencent Robotics X, Shenzhen, China and Shenzhen Unity Drive, Shenzhen, China, since 2019. He is the inventor of the world's first ultra low-cost circuit for 3-D and higher dimensional registration. He is currently in charge of some open projects of robotics and aerospace engineering from several State Key Laboratories in China.
\end{IEEEbiography}

\end{document}